# Electrochemical Doping in Ordered and Disordered Domains of Conjugated Polymers

Priscila Cavassin, Isabelle Holzer, Demetra Tsokkou, Oliver Bardagot, Julien Réhault, Natalie Banerji*

P. Cavassin, I. Holzer, Dr. D. Tsokkou, Dr. O. Bardagot, Dr. J. Réhault, Prof. N. Banerji

Department of Chemistry, Biochemistry and Pharmaceutical Sciences, University of Bern, Freiestrasse 3, 3012 Bern, Switzerland

E-mail: natalie.banerji@unibe.ch

Conjugated polymers are increasingly used as organic mixed ionic-electronic conductors in electrochemical devices for neuromorphic computing, bioelectronics and energy harvesting. The design of efficient applications relies on high electrochemical doping levels, high electronic conductivity, fast doping/dedoping kinetics and high ionic uptake. In this work, we establish structure-property relations and demonstrate how these parameters can be modulated by the co-existence of order and disorder. We use *in-situ* time-resolved spectroelectrochemistry, resonant Raman and terahertz conductivity measurements to investigate the electrochemical doping in the different morphological domains of poly(3-hexylthiophene). Our main finding is that bipolarons are found preferentially in disordered polymer regions, where they are formed faster and are thermodynamically more favoured. On the other hand, polarons show a preference for ordered domains, leading to drastically different bipolaron/polaron ratios and doping/dedoping dynamics in the distinct regions. We evidence a significant enhancement of the electronic conductivity when bipolarons start being formed in the disordered regions, while the presence of bipolarons in the ordered regions is detrimental for transport. Our study provides significant advances in the understanding of the impact of morphology on the electrochemical doping of conjugated polymers and the induced increase in conductivity.

## 1. Introduction

Electrochemical doping is one of the most efficient and controlled ways to tune the conductivity of organic semiconductors by several orders of magnitude.[1-3] Because this doping process is reversible, it allows precise modulation of the conductivity level, making it an important tool for a wide range of novel applications, such as neuromorphic systems,[4,5] thermoelectric generators,[6,7] batteries,[8,9] and organic electrochemical transistors (OECTs) for biological interfacing.[10-12] In electrochemical doping, electronic charge carriers are injected into the organic thin film from an electrode, while counterions from an electrolyte infiltrate the film to maintain charge neutrality. The magnitude of the voltage applied between the film and a counter electrode tunes the conductivity by controlling the density of charge carriers in the film.[13] When the film becomes positively charged, this is referred to as p-type doping or alternatively as oxidation with simultaneous compensation by negative ions.

Ideal organic electrochemical devices require fast doping and depoping kinetics (*i.e.,* fast ON/OFF switching) combined with high ionic uptake (*i.e.,* high capacitance), and high conductivity (*i.e.,* high doping levels and high mobility of the electronic carriers). Both the intrinsic material properties and the morphology of the film need to be fine-tuned to simultaneously optimize these parameters. This requires clear structure-property relations that are currently not well understood. Here, we investigate electrochemical doping of the archetypal conjugated polymer poly(3-hexylthiophene) (P3HT) by *in-situ* spectroelectrochemical techniques probed in the visible-near-infrared (Vis/NIR), Raman and THz range. We correlate the generation dynamics of different charged species (polarons, bipolarons) in distinct morphological domains to the nanoscale conductivity at different doping levels.

Most conjugated polymers are indeed semi-crystalline, with a mixture of ordered and disordered domains in thin films. This heterogeneity impacts the doping process and considerably complicates its characterization, as highlighted for P3HT in recent X-ray diffraction studies.[14-20] For instance, Guardado *et al*. and Thomas *et al*. have independently demonstrated that at low electrochemical doping levels, ions are preferentially located in the disordered domains of P3HT despite polarons being formed in the more ordered domains. [19,21] It is only at high doping levels that ions also penetrate the ordered domains. A drawback of these studies is that X-ray diffraction accesses only the crystalline domains of the film. Alternatively, *in-situ* electrochemical resonant Raman spectroscopy (ERRS) is sensitive to subtle changes in the polymer backbone conformation, so that it can distinguish neutral and charged species in the ordered and disordered regions. Nightingale *et al.* have used this technique to show that the ordered polymer domains undergo more pronounced changes in molecular conformation when they are doped.[20] However, bipolarons were not considered in this study and are often ignored in electrochemical investigations of P3HT.[14,18,19,21] They are nevertheless present at high doping levels and their Raman signatures were recently discussed by Mansour *et al.* in molecularly doped P3HT.[22] We, as well as Neusser *et al.*, have demonstrated that those bipolarons play an important role in the conductivity of the electrochemically doped polymer.[17,23] Now, further understanding on how the presence of polarons and bipolarons in different morphological domains impacts the transport is essential, and therefore it is critical to differentiate between the doping mechanisms that take place in each region of P3HT films.

In this work, we offer novel insights on: i) the generation of polarons and bipolarons in the distinct domains, ii) the doping of the disordered domains, which has been much less explored, iii) the kinetics of the redox reactions involved at various doping and dedoping levels, and iv) the impact of having different carriers in the ordered and disordered regions on the short-range transport properties. We find that in contrast to polarons, bipolarons are preferentially found within disordered domains, which by themselves have low conductivity. Nevertheless, the high charge density stored as bipolarons in these regions is beneficial to the overall transport up to a critical bipolaron concentration. We demonstrate that all aspects of electrochemical doping in conjugated polymers (ionic and electronic doping level, degree of oxidation, doping and dedoping kinetics, conductivity) differ significantly in the ordered and disordered regions of the film and can be controlled by tuning their co-existence. This provides opportunity for materials design and morphological optimization.

## 2. Results and discussion

### 2.1 Evolution of the redox species at different voltages

We first measure the absorbance changes of a P3HT thin film (~ 20 nm) immersed in aqueous $KPF_6$ electrolyte (0.1 mol $L^{-1}$) at different electrochemical doping potentials. Efficient electrochemical doping of P3HT in this electrolyte has been previously reported.[24] For those spectroelectrochemistry measurements, we use a home-built setup that applies a square voltage pulse and simultaneously measures the current response and the visible to near infrared (Vis/NIR) absorbance of the film (Figure 1a). The films are spin-coated on a conductive ITO substrate from chloroform solution, yielding a relatively disordered morphology.[25] The voltage is applied on the Ag/AgCl electrode ranging from +0.2 V to -0.8 V, with steps of 0.1 V. Before each measurement, five doping/dedoping cycles are performed to stabilize the response. In Figure 1b, the evolution of the spectra at different voltages is displayed. Initially, the absorbance is centered around 500 nm, but as the applied voltage becomes more negative, we observe a decrease of this band concomitant with the rise of a band around 800 nm, followed by the rise of another band around 1400 nm at the most negative voltages. To deconvolute the overlapping absorbance signatures and their evolution with the applied voltage, we use Multivariate Curve Resolution (MCR) analysis (Supplementary Note 1). We find that the absorption spectra can be decomposed into four different spectral signatures (Figure 1c). The data reconstruction from the spectral components (MCR fit) is shown in Supplementary Fig. 1. Their absorbance cross section is estimated by relating the absorbance to the density of injected charge, measured by chronoamperometry on films of different thicknesses (Supplementary Note 2, Supplementary Fig. 2, Supplementary Table 1).

The absorption around 500 nm is attributed to P3HT in the neutral state. The MCR analysis decomposes this neutral band into two distinct spectral signatures (Figure 1c): i) The neutral polymer in more disordered regions (amorphous phase), which has a broad Gaussian-like peak centered at 495 nm, and ii) the polymer in more ordered domains of the film (crystalline phase), that exhibits the two characteristic 0-0 and 0-1 vibronic bands centered at 560 nm and 605 nm. Both spectral signatures are similar to the absorption spectra of disordered and ordered P3HT that have been directly measured and reported in the literature.[26,27] Also in agreement with previous studies, the spectral components at 800 nm and 1400 nm are assigned to the charged species, respectively to polarons and bipolarons.[16,28] Upon oxidation of P3HT, a single electron is removed from the polymer chain and a localized lattice distortion occurs. Together, they are defined as a positive polaron.[16,29] When another electron is removed from the P3HT chain, both polarons can be stabilized by sharing the same local lattice distortion, resulting in the formation of a bipolaron.[29] Enengl *et al.* have experimentally demonstrated that the charged species present in electrochemically doped P3HT are polarons and bipolarons, excluding the presence of polaron pairs.[16] This is in agreement with recent electron paramagnetic resonance (EPR) data on such films.[17] Modifications to the electronic energy levels due to the presence of polarons and bipolarons lead to the new absorbance bands at 800 nm and 1400 nm respectively, seen by spectroelectrochemistry.[16,17,29-31]

Figure 1d shows how the density of each redox species (obtained from the MCR concentrations and absorbance coefficients) evolves with applied voltage. We distinguish four different doping regimes. Between +0.2 V and 0 V (regime 0), there is no significant change in the spectra, which indicates an oxidation onset at 0 V in aqueous 0.1 mol $L^{-1}$ $KPF_6$ electrolyte. In regime 0, mainly neutral species are present, whereby our analysis shows that the film is composed of about 57% disordered and 43% ordered sites, in accordance with literature for similar molecular weight and coating conditions.[32] Note that the electrochemical pre-cycling of the film results in a non-zero density of polarons below the oxidation onset, likely because a few ions are trapped inside the film.[33] From 0 V to -0.2 V (regime I), there is a decay of both ordered and disordered neutral species concomitant with the rise of the polaronic population. Between -0.3 V to -0.5 V (regime II), both the populations of polarons and bipolarons are increasing while the neutral signatures further decrease. Finally, for voltages more negative than -0.5 V (regime III), the ordered neutral population is completely depleted, while the disordered neutral population continues to decay. This suggests a lower oxidation potential for the neutral ordered species (crystalline phase) compared to the amorphous phase, as previously reported.[19,21,34,35] The fact that the distinct morphological domains dope at different voltages leads to the apparent blue-shift of the absorption band around 500 nm (Figure 1b).[16,19] Moreover, in regime III, the bipolaron density continues to increase, while the polaron density decreases, suggesting that part of the polarons is converted to bipolarons.[16] At -0.8 V, the film is completely doped and contains predominantly polarons (47%) and bipolarons (53%).

Summing the density of all species shows that the total density increases at the negative doping voltages, in particular when bipolarons are being formed (Figure 1d). The neutral species correspond to chromophoric units (ground state polymer segments responsible for light absorption), while the charged species are structurally modified polymer segments from which one or two electrons have been removed. Based on the reported film density of P3HT,[36] the density of monomers is around $4x10^{21}$ $cm^{-3}$. Relating this to the total density of neutral species at +0.2 V (5.3 $x10^{20}$ $cm^{-3}$, Figure 1d), we find that the chromophores extend roughly over 7-8 monomers. Since the total density stays constant when neutral species are converted to polarons (regime I), we conclude that the polarons have a similar spatial extent as the neutral chromophores. However, at -0.8 V about twice the density of bipolarons is formed compared to what is expected from a one-to-one conversion from polarons, implying that bipolarons are more localized and span only 4 monomers.

**2.2 Structural backbone changes during electrochemical doping**

Although some spectral changes for localized and delocalized charges of doped P3HT in solution have previously been reported,[37] the broadness of the polaron and bipolaron bands in our Vis/NIR spectroelectrochemistry film data does not allow to distinguish whether those species are present in the ordered or disordered domains of P3HT. Therefore, we switch to *in-situ* electrochemical resonant Raman spectroscopy (ERRS), which is more sensitive to the structure and environment of the polymer backbone.[20] For experimental reasons, the P3HT film (thickness of approximately 40 nm) is now spin-coated on two short-circuited gold electrodes separated by a 2 mm gap (‘OECT-like’ device), unlike the planar ITO-based sample of the previous section. The device is immersed in the electrolyte together with the Ag/AgCl electrode and the Raman spectra are acquired after reaching the steady state for voltages ranging from +0.2 V to -0.8 V with steps of 0.1 V. The excitation wavelength used is 633 nm, which is mainly resonant with the ordered neutral absorption band of P3HT. Thus, the Raman intensity strongly decays as ordered P3HT is depleted, so that the spectra are normalized in order to compare the changes upon doping (Supplementary Fig. 3).

The normalized Raman spectra at selected voltages are shown in Figure 2a. We assign the observed bands based on previous Raman studies on doped and undoped P3HT.[20,22,38,39] When the film is undoped (V > 0 V), the Raman spectra show an intense band at 1455 $cm^{-1}$ and a weaker band at 1380 $cm^{-1}$, attributed to the $C_\alpha - C_\beta$ and to the $C_\beta - C_\beta$ stretching vibrations, respectively (see inset of Figure 2a).[20,38,39] When the film is doped and predominantly polarons are formed (-0.2 V to -0.5 V), the $C_\alpha - C_\beta$ vibrational mode shifts towards smaller wavenumbers to the 1405-1438 $cm^{-1}$ region.[20,38,39] This shift is due to a more planar backbone and more delocalized π-electrons in the polaronic state. Nightingale *et al*. have combined DFT calculations and ERRS to demonstrate that the $C_\alpha - C_\beta$ bands of neutral and polaronic P3HT consist of a convolution of signals from ordered and disordered domains.[20] For neutral P3HT, disordered chains vibrate at higher frequency than more ordered chains due to a shorter conjugation length. Conversely, polaronic disordered chains vibrate at a lower frequency than ordered ones, since the ordered segments undergo weaker lattice distortions compared to the large conformational changes when polarons form in disordered P3HT.[20] When we go to even higher doping levels where bipolarons are generated (< -0.5 V), two peaks appear at 1465 and 1493 $cm^{-1}$. Mansour *et al.* have chemically doped P3HT and observed bipolaronic peaks at similar wavelengths.[22] They have assigned the peak at ~1460 $cm^{-1}$ to the $C_\alpha - C_\beta$ symmetric stretching mode of disordered bipolarons, and the one at ~1500 $cm^{-1}$ to the $C_\alpha - C_\beta$ asymmetric stretching, which appears when high disorder is induced at high chemical doping levels. Unlike for chemical doping, the electrochemical oxidation of P3HT does not necessarily lead to enhanced disorder. Moreover, in our data, we observe that the peak at 1493 $cm^{-1}$ appears only in regime III, when ordered polarons are converted to bipolarons. Therefore, we rather assign this band to the formation of bipolarons in the ordered domains. The blue-shift of both bipolaron bands has been previously attributed to a shorter conjugation length,[22] in full agreement with our observation that the bipolarons are about twice as spatially localized as the neutral and polaron sites.

Following the procedure of Nightingale *et al.*,[20] we use Gaussians to deconvolute the signature of the $C_\alpha - C_\beta$ stretching mode into contributions of ordered and disordered species in different redox states (Figure 2a, Supplementary Fig. 4 and Supplementary Table 2). As seen before with Vis/NIR spectroelectrochemistry, some polarons are already present at 0 V (regime 0) and this effect is slightly enhanced in the OECT-like configuration. The ERRS analysis now allows to assign those trapped polarons to the ordered regions of the film. At -0.5 V (regime II), a strong decrease of both the ordered and disordered neutral signatures, an increase of the polarons in both regions and a rise of the disordered bipolarons are observed. Finally, at -0.8 V (regime III), the signal from polarons in the ordered domains decreases while the ones from the disordered polarons and bipolarons in both domains rise. We note the presence of neutral segments even at the highest doping voltages, in contrast to the data in Figure 1. This could be related to less efficient doping in the OECT-like geometry or to enhancement of the neutral bands due to their resonance at 633 nm.[40]

The area of the deconvoluted peaks give us insights about the evolution of the species upon doping. Figure 2b depicts the relative peak area (normalized by the total peak area summed over all Gaussians) for the different redox species as a function of the applied voltage. The data corroborates that the ordered neutral regions are oxidized at slightly lower potential than the disordered ones. Additionally, ERRS demonstrates that polaron formation in the ordered domains occurs before that in the disordered domains. The opposite trend is seen for the bipolarons, which form in the disordered domains at less negative voltages. Although we did not correct for the oscillator strength of the Raman bands, it is remarkable that their sum reproduces extremely well the evolution of the different neutral and charged species measured by Vis/NIR spectroscopy (measured on the same OECT-like device as the ERRS, Figure 2b and Supplementary Fig. 5). This validates our interpretation for both spectroscopic techniques and confirms the observed trends. An important conclusion from ERRS is that charge carriers remain in both the ordered and disordered domains once steady-state conditions are reached (in the absence of applied source-drain voltage). In contrast, in organic solar cell blends, photogenerated electrons migrate to more ordered fullerene clusters that act as energetic sinks.[41] Also, it has been shown that transport in P3HT occurs predominantly in the ordered regions and along tie chains linking them together.[42] Here, we cannot exclude a dynamic equilibrium between the charges in different morphological regions, but the overall charge density of polarons and bipolarons stays in the region where it was initially generated.

**2.3 Dynamics of the electrochemical reactions**

Time-resolved spectroelectrochemistry in the Vis/NIR range allows to go beyond the steady-state response and to monitor the dynamics of the electrochemical doping. This is key to understand the mechanisms of the redox processes and how they differ for the distinct morphological domains. Using the same experimental setup shown in Figure 1a and the planar ITO-based device, we measure the spectral response to a square voltage pulse with a time resolution of 10 ms, much higher than what can be achieved by state-of-the-art X-Ray scattering techniques.[15] Each voltage pulse occurs from +0.4 V (dedoping voltage) to a doping voltage that is stepped down from -0.1 V to -0.8 V and back to +0.4 V (Supplementary Fig. 6). Supplementary Fig. 7 shows the dynamics for P3HT films with 4 different thicknesses (5, 20, 55 and 110 nm) for a doping voltage of -0.8 V. The dynamics of the thicker films gradually slow down since they are kinetically limited by ion diffusion,[43] while the dynamics converge to a similar time evolution for the thinner films of 20 nm and 5 nm thickness. In this range, ion diffusion is no longer the limiting factor, and the intrinsic rates of the redox reactions are observed. These are in the focus of the current work, therefore a thin film of 20 nm is used in the following analysis.

The temporal evolution of the density of disordered neutral, ordered neutral, polaron and bipolaron species (from MCR decomposition with the components in Figure 1c) is depicted in Figure 3 for representative doping and dedoping processes in regime I (-0.2 V), regime II (-0.5 V) and regime III (-0.8 V). For a first assessment of the data, we estimate the rate constants of the electrochemical doping and dedoping processes by using a sum of exponentials to fit the dynamics (Supplementary Notes 3-4, Supplementary Figures 8-11, Supplementary Tables 3-4). Typically, we require two exponentials to describe the conversion between redox species, as well as offsets (equal to the steady-state densities) that represent the equilibria establishing between the species at each voltage. The faster component usually accounts for about 80-90% of the conversion. The biexponential behaviour can be explained by some structural and energetic inhomogeneity of the polythiophene chains in the ordered and disordered regions.[44] Moreover, the doping dynamics at the earliest times cannot always be reproduced with the biexponential function, especially at low doping voltages. We tentatively ascribe this short delay to ions reaching the proximity of the electrochemical reaction site and discard it during the exponential fitting. The effect becomes negligible for the dedoping dynamics.

The kinetic traces in Figure 3 and their exponential analysis reveal clear correlations between the dynamics of the neutral and charged species, allowing to deduce how they interconvert. During doping, the decay of ordered neutral species correlates mainly with the formation of polarons and is slightly faster (at a given voltage) than the decay of disordered neutral species, which is concomitant with the rise of bipolarons below -0.3 V (Supplementary Table 3). This agrees with the lower oxidation potential of the ordered domains and the preferential formation of bipolarons in amorphous regions.[19,20] As expected for electrochemical reactions, the rates increase at increasing doping voltage. For example, the average time constant for the neutral ordered decay evolves from $\tau$ = 1.0 s (at -0.3 V) to $\tau$ = 0.2 s (at -0.8 V), and the neutral disordered decay accelerates from $\tau$ = 2.3 s to $\tau$ = 0.9 s within this voltage range. Moreover, the final equilibria shift towards a higher doping level at more negative voltages, with the offsets of the neutral species decreasing and the offsets of the polarons and bipolarons increasing. For the highest doping voltages (-0.6 V to -0.8 V), the polaron population first peaks and then decays with a time constant of $\tau \approx$ 5-6 s that is correlated with the slower bipolaron rise. We assign this to the conversion of polarons to bipolarons in the ordered regions (see kinetic model below).

For the dedoping, the switch always occurs to a voltage of +0.4 V, so that the final population (dominated by the neutral species) is independent of the initial doping voltage (Figure 3, Supplementary Table 4). This nevertheless determines the doping level at the start of the process. We find that dedoping is generally faster than doping and shows a much weaker dependence on the voltage jump. Essentially the inverse reactions occur than during doping. For all regimes, the doping is completely reversible, which indicates that no permanent morphological disruption is induced by the injection of ions in our experimental conditions and after the pre-cycling.[21] When initially present, the bipolarons show a fast decay (τ = 0.1 s from -0.3 V to -0.6 V), which slightly slows down when more bipolarons are present in the ordered domains (τ = 0.2 s from -0.7 V to -0.8 V). At the most negative voltages, a concomitant rise of the polarons is observed. Apart from that, the polarons decay with an average time constant that ranges from 0.4 s to 0.7 s between -0.1 V and -0.8 V, and which correlates to the rise of the disordered neutral species. We note that even the fastest doping and dedoping dynamics measured here for P3HT are at least one order of magnitude slower than the ones we recently reported for the electrochemical reactions of PEDOT:PSS (for a step from +0.1 V to -0.6 V and back).[45]

**2.4 Kinetic modelling of the reaction mechanisms**

Based on these observations, we build a kinetic model to describe the doping and dedoping mechanisms that occur in the different regions of P3HT (Supplementary Notes 5-6). Such a model confirms the reaction sequence and additionally disentangles the dynamics of the charges in the ordered and disordered regions. The model is implemented using the symfit Python package.[46] To simplify, we discard the initial non-exponential delay in the doping dynamics at low voltages, and we account only for the predominant fast component of the conversion between the redox species (ignoring the ~10-20% slower component due to inhomogeneity). As closure is required for kinetic modelling, we divide the bipolaron population by 1.37 during the fitting to keep the total species density constant. We know from the Raman data that distinct polarons and bipolarons in the ordered and disordered regions are formed, so that their population is subdivided into ordered and disordered species in the model. The sum is fit to the experimental dynamics. No exchange between the species in the two domains is necessary to obtain a good fit, so we assume that charges stay in the region where they are generated. Finally, to account for the equilibria between redox species at each doping voltage, we introduce back-transfer rates for all processes. The fits to the kinetic model are shown in Figure 3 and Supplementary Figs. 12-17. The fit parameters are summarized in Tables 1-2 for the doping and dedoping, respectively, and are depicted as a function of the applied voltage in Figure 4.

**Regime I (-0.1 V to -0.2 V)**. Here, neutral species in both ordered and disordered domains are converted to polarons, resulting in equilibria ($N_{dis} \rightleftarrows P_{dis}$ and $N_{ord} \rightleftarrows$ $P_{ord}$) between the redox states. At increasing doping voltages, the rate of polaron generation increases and the equilibria shift towards their formation. It has previously been reported using X-ray diffraction, that at such low doping voltages, the electrolyte anions mainly enter the amorphous regions of P3HT, but most of the polarons are formed in the more ordered polymer regions (at the interface to disordered domains, see schematics in Figure 3a).[19,21] Our analysis allows to quantitatively determine that 66% of the polarons are formed within the ordered regions at -0.2 V. Also, we find that the conversion to polarons is about three times faster in the ordered regions and the higher equilibrium constant points to a more favourable (more exergonic) oxidation. When switching back to +0.4 V from either -0.1 V or -0.2 V, the reduction to the neutral state is 1-2 orders of magnitude faster than the initial oxidation, occurs on a similar time scale in the ordered and disordered regions and shows a weaker dependence on the voltage jump. The conversion to neutral species ($P_{dis} \rightleftarrows N_{dis}$ and $P_{ord} \rightleftarrows N_{ord}$) is almost complete, except for a small concentration of polarons remaining in the ordered regions, likely due to trapped ions and lattice changes during the initial pre-cycling.

**Regime II (-0.3 V to -0.5 V).** The trends from the previous regime continue, with the conversion to polarons becoming faster and more favourable with increasing doping voltage in both morphological regions, and the doping being generally faster in the ordered domains. The total charge density now reaches $4 \times 10^{20}$ $cm^{-3}$, which has been reported as the threshold for ion penetration into the ordered regions of P3HT, leading to important crystalline changes and enhanced mobility.[19,21] At -0.5 V, about 66% of all polarons are formed within the ordered regions and 94% of the ordered neutral density decays according to $N_{ord} \rightleftarrows P_{ord}$ (Figure 3b). In contrast, only 66% of the neutral disordered segments are converted to charges. Additionally, there is conversion of polarons to bipolarons (increasingly enhanced at more negative voltages), but this occurs only in the disordered domains ($N_{dis} \rightleftarrows P_{dis} \rightleftarrows B_{dis}$). The rate of this process is comparable (slightly slower) to the preceding conversion of disordered neutral segments to polarons, so that the disordered neutral decay appears to correlate to the bipolarons rise in the experimental data. In agreement with preferential bipolaron formation in the disordered regions, Nightingale *et al*. suggest that the oxidation peak of the amorphous phase coincides with bipolaron formation.[20] This is promoted by the higher reorganization energy and charge localization found in disordered P3HT domains,[20] as well as by stabilization of the bipolarons by the ions present predominantly in these regions. Switching back to a dedoping voltage of +0.4 V in regime II leads to very fast and complete back-conversion in the disordered regions according to $B_{dis} \rightarrow P_{dis} \rightleftarrows N_{dis}$. In the ordered domains, the reduction of polarons ($P_{ord} \rightleftarrows N_{ord}$) is slightly slower and slows down with increasing voltage jump.

**Regime III (-0.6 V to -0.8 V).** At the highest doping voltages, all the oxidation processes in the ordered and disordered regions become even faster and more favourable. The conversion of neutral species to charges is over 97% complete at -0.8 V, even in the disordered domains, where similar densities of polarons and bipolarons are formed according to $N_{dis} \rightleftarrows P_{dis} \rightleftarrows B_{dis}$. About 20% of all bipolarons are now also generated in the ordered regions ($N_{ord} \rightarrow P_{ord} \rightleftarrows B_{ord}$), but at a much slower rate, leading to a relatively low equilibrium concentration. It is noteworthy that this conversion in the ordered domains only starts after all neutral species have been depleted by at least 97%, resulting in a peak of the polaron population before they slowly decay to bipolarons. We have recently reported a similar behaviour in PEDOT:PSS.[45] The penetration of ions into the ordered domains in regime III might be required for bipolarons formation. Upon dedoping (to + 0.4 V), the ordered bipolarons decay about four times faster than the disordered bipolarons, while the conversion of polarons to neutral species is faster in the disordered domains.

The overall evolution of the rate and equilibrium constants with the doping and dedoping voltages is summarized in Figure 4. The faster conversion of neutral species to polarons in the ordered regions, as opposed to the much faster conversion of polarons to bipolarons in the disordered regions, is clearly seen during doping (Figure 4a). For dedoping, the rates follow the opposite trend and are highest for the bipolaron to polaron conversion in the ordered regions, while the subsequent formation of ordered neutral species is slowest (Figure 4b). The doping rates (shown on a log scale) show an increase and gradual saturation as the oxidation voltage becomes higher, whereas the dedoping rates slightly decrease when the initial voltage is more negative (except for the ordered polaron to neutral conversion, which first increases and then decreases). We tentatively ascribe this behaviour to the impact of increasing driving force in the Marcus normal and inverted region, respectively, with a more detailed analysis planned in subsequent work. Finally, $\ln(K)$ for all doping processes increases linearly with voltage (Figure 4c), as expected according to Nernst equation. The equilibrium constants clearly highlight that polarons are thermodynamically more favoured in the ordered regions, while bipolarons prefer the disordered ones.

### 2.5 Impact of film morphology on the redox processes

To verify the impact of the ordered and disordered domains on the electrochemical doping of P3HT, two additional samples with slightly higher order (regioregular P3HT spin fast from o-dichlorobenzene solution) and with much lower order (regiorandom RRa-P3HT cast from chloroform) are investigated. The absorption spectra of the dry films in Supplementary Fig. 18 show that the 0-0 vibronic shoulder is enhanced and the spectrum red-shifted in the regioregular film cast from o-dichlorobenzene compared to the one cast from chloroform. On the other hand, the RRa-P3HT film is essentially amorphous, as confirmed by the absence of any vibrational structure. Moreover, the absorbance band of RRa-P3HT is notably blue-shifted, which is attributed to a shorter conjugation length, higher disorder and reduced intermolecular interactions.[47]

The steady-state spectroelectrochemical data obtained in the Vis/NIR range for the regioregular P3HT films cast from o-dichlorobenzene and chloroform is compared in Figure 5a and Supplementary Fig. 19 (measured in OECT-like configuration on 35 nm thick films in 0.1 M aqueous $KPF_6$ electrolyte). The same MCR components as before are used for the analysis. The overall evolution of the redox species with voltage is comparable in the two samples, but some clear differences are seen. First, there are more ordered compared to disordered neutral chromophores initially present in the film cast from o-dichlorobenzene, as expected for more crystalline P3HT. Moreover, more polarons and less bipolarons and are formed upon doping this film, with a bipolaron/polaron ratio of 0.70 at -0.8 V compared to 0.97 for the more disordered film cast from chloroform. This confirms preferential bipolaron formation in disordered regions of P3HT and shows that the bipolaron/polaron ratio in the doped state can be controlled by tuning the morphology. Finally, we notice a high initial polaron concentration at positive voltages, which is related to the experimental conditions (OECT-like device, shorter dwell time). The effect is more pronounced in the chloroform sample, possibly because trapping of ions is enhanced in the more amorphous film.

The absorbance at different voltages for the RRa-P3HT sample (Figure 5b) is recoded on a 125 nm thick film spin-coated on an ITO substrate and doped with 0.1 M $KPF_6$ in acetonitrile electrolyte. Unlike regioregular P3HT, RRa-P3HT cannot be doped using aqueous $KPF_6$. Because of the electrolyte change, the electrochemical window is shifted to a doping threshold at -0.7 V *vs* Ag/AgCl and a maximum doping level at -1.4 V. Within the investigated window, we never reach a regime where all neutral species are oxidized and there is no significant drop in polaron population, which reaches a constant value while the bipolarons still keep rising (Supplementary Fig. 20). Significantly different MCR signatures are found for RRa-P3HT (Figure 5c). Only disordered and no ordered neutral species are present, with a very blue-shifted band at 400 nm. The polaron and bipolaron spectra are also blue-shifted due to the enhanced disorder, so that two polaronic transitions can now be seen within the spectral window. The blue shift might also indicate that the $PF_6^-$ anions are located closer to the polymer backbones in amorphous P3HT.[48] The time-resolved optical response shown in Figure 5d is measured while applying a voltage of -1.2 V to the RRa-P3HT film. The neutral state decays with a concomitant rise of the polarons and a slower rise of the bipolarons (Supplementary Note 6, Supplementary Fig. 21, Supplementary Table 5). The conversions can be kinetically modelled using $N_{dis} \rightleftarrows P_{dis} \rightleftarrows B_{dis}$ (Supplementary Fig. 22, Supplementary Table 6). Although the film is thicker, we do not expect ion diffusion to become a limiting factor, since this is very efficient in amorphous P3HT. The rate and equilibrium constants are $k_{N,dis \rightarrow P,dis} = 1.94\, s^{-1}$ (with $K_{N,dis \rightleftharpoons P,dis} = 1.1$) and $k_{P,dis \rightarrow B,dis} = 0.36\, s^{-1}$ (with $K_{P,dis \rightleftharpoons B,dis} = 0.3$). The dynamics is similar to the disordered regions of regioregular P3HT for a similar overpotential (- 0.6 V), also depicted in Figure 5d. This shows that we reach here the intrinsic electrochemical kinetics of amorphous P3HT.

### 2.6 Short-range conductivity of electrochemically doped P3HT

We have previously shown using terahertz (THz) spectroscopy that the coexistence of polarons and bipolarons is essential to maximize the short-range conductivity of electrochemically doped P3HT.[23] The conductivity was found to first increase with bipolaron concentration and then to plateau or to decrease. Similar findings for the macroscopic conductivity were reported by Neusser *et al.* for the same system.[17,23] The behavior was attributed to mixed valence conduction where transport occurs by hole hopping between polaronic and bipolaronic sites.[49] In addition, we can now evaluate the effect of having bipolarons in the ordered and disordered regions. The impact of charges in different morphological domains on the conductivity of electrochemically doped P3HT has not been previously addressed.

We revisit here data that we have initially presented as part of a THz study on P3HT.[23,50] To obtain a measurable conductivity during electrochemical oxidation, without obstruction of the THz beam by the electrodes, a thicker polymer film (270 nm cast from 1,2-dichlorobenzene) and an OECT-like sample configuration with a 4 mm gap between the gold electrodes is used. THz and Vis/NIR data recorded in 0.1 M $TBAPF_6$/acetonitrile electrolyte is presented, where a higher doping level can be reached (Figure 6). The MCR analysis is carried out with the components and absorption cross sections used throughout this manuscript (Figure 6a). Apart from a shift of the onset of bipolaron formation to about -0.8 V in acetonitrile (instead of -0.2 V in water), a similar evolution of the redox species with voltage is obtained as in Figure 1d, allowing to assign the different doping regimes. The short-range conductivity at 1 THz starts to be detectable in regime II, when bipolarons are being formed in the disordered regions, and continues to rise into regime III, where the bipolaron concentration increases in both the disordered and ordered domains (Figure 6a). However, at -1.4 V, the conductivity saturates and then drops.

The conductivity depends on the effective THz mobility and the conductive charge density, which are disentangled using the Drude-Smith model as previously described (Figure 6a).[23] The mobility shows a weak initial rise, but then stays constant and does not correlate with the evolution of the conductivity. On the other hand, the conductive charge density follows a similar trend as the conductivity and drops at the highest oxidation voltages. The linear correlation of the conductivity to the density of conductive charge confirms the predominant role of charges contributing to transport (Figure 6b). The conductivity decreases at high doping levels although more charges keep being injected, because the fraction of conductive charge with respect to injected charge drops from 42% to 27%. In the investigated voltage range, charge injection leads predominantly to bipolaron formation. The fraction of charges in bipolarons (= 2B/(2B+P), considering two charges per bipolaron) rises linearly with the injected charge density to 71% (Figure 6c). Relating the fraction of conductive charge to the fraction of bipolaronic charge identifies two trends (Figure 6d). First, a linear rise with a slope above 1, showing that every injected charge (converting a polaron to a bipolaron) adds about one conductive charge to the transport. Second, when the bipolaronic fraction reaches 45%, the trend reverses. Adding more bipolarons now linearly decreases the fraction of conductive charge with a slope of -0.8.

We know from the present study that bipolarons prefer to form in the disordered regions of P3HT in regime II and early regime III when the rise in conductivity is observed. We cannot detect any short-range THz conductivity for electrochemically doped RRa-P3HT in spite of a high bipolaron density, suggesting that THz spectroscopy probes the transport in the ordered regions for the regioregular sample. The macroscopic conductivity in RRa-P3HT is also at least one order of magnitude lower than in the regioregular polymer.[17,51] Indeed, the disordered domains are considered the limiting factor for macroscopic transport, which occurs through ordered domains that are interconnected across disordered regions via tie chains.[32,52,53] Nevertheless, we find that the conductivity in the ordered regions is impacted by bipolaron formation in the disordered domains. At the nanoscale, this is likely an interfacial effect, whereby the transport between the ordered and disordered regions is facilitated by holes hopping from disordered bipolarons to ordered polarons, allowing more charges to participate in the short-range transport. The bipolarons might then dissociate to polarons in the ordered regions where they are less stabilized, explaining why we do not observe net bipolaron migration between the domains. At the macroscopic scale, the better transport across the ordered/disordered interface is extended along the tie chains that connect the ordered domains. Having a high density (due to their localized nature) and being doubly charged, the bipolarons in the disordered regions represent an important reservoir of charge that favors transport along doped tie chains.

In regime III, the conductivity decreases above a certain bipolaron threshold, which we relate to two effects. First, the concentration of bipolarons in the disordered regions starts to exceed the one of polarons at high oxidation voltages (Supplementary Figs.

13-14), which is unfavorable for mixed valence conduction and causes hole-hole repulsions.[54] Second, bipolarons are being formed in the ordered domains, which strongly localizes the charges to a given site. They are stabilized by ions entering the ordered regions, which can cause morphological disruption. The rise of ordered bipolarons at the highest doping levels can therefore also decrease the conductivity.

**3. Conclusion**

Organic mixed ionic-electronic conductors constitute a variety of electrochemical devices used in neuromorphic computing, bioelectronics and energy harvesting. Their efficient application critically depends on high electrochemical doping levels with mobile electronic charge carriers, fast doping/dedoping kinetics and high ionic uptake. We establish here structure-property relations and show how these parameters can be controlled by the co-existence of ordered and disordered domains in conjugated polymers. We apply a variety of *in-situ* spectroscopic techniques to functioning electrochemical devices based on the prototypical P3HT system, including time-resolved spectroelectrochemistry, resonant Raman experiments and terahertz conductivity measurements.

Our main finding is that bipolarons are found preferentially in disordered polymer regions, where they are formed faster and are thermodynamically more favoured. We relate this to their better stabilization at high ionic content in amorphous P3HT domains, and to higher energetic gain when forming a bipolaron on an initially disordered and localized polymer segment. We provide the following experimental evidence for the effect: i) The polaron-to-bipolaron oxidation onset occurs at lower doping voltage in the disordered compared to the ordered regions (by ~0.2 V). ii) The steady-state bipolaron/polaron equilibrium is more shifted towards bipolarons in disordered domains. iii) The rate of disordered bipolaron formation during doping is about two orders of magnitude higher than for ordered bipolarons, while the rate of their depletion during dedoping is three times slower. In contrast to bipolarons, polarons show a (less pronounced) preference for the ordered domains, where the neutral-to-polaron oxidation onset is lower, both polaron/neutral and bipolaron/polaron equilibria are shifted more towards polarons, and the polaron formation is faster. Given the opposite preference of polarons and bipolarons for ordered and disordered regions, respectively, we demonstrate that the bipolaron/polaron ratio can be tuned via the morphology in films with different degrees of order.

Amorphous polymer domains are known to favour ionic uptake. Additionally, we show here that they promote the formation of bipolarons, which constitute an important reservoir of electronic charge in the disordered regions. Those bipolarons have a significant impact on transport. An important enhancement of the electronic conductivity occurs when bipolarons start being formed in the disordered regions, caused by an increase of the conductive charge fraction. Since the disordered regions are by themselves not very conductive, we stipulate that the high charge density in the disordered domains helps interfacial transport across ordered/disordered boundaries and along tie chains that connect the ordered domains. At high doping levels, the trend reverses and the conductivity decreases once the density of bipolarons exceeds the one of polarons in the disordered regions (unfavourable for mixed valence conduction), and once bipolarons and ions are present in the ordered regions. In conclusion, all aspects of electrochemical doping in conjugated polymers (ionic and electronic doping level, degree of oxidation, doping and dedoping kinetics, conductivity) differ significantly in the ordered and disordered regions of the film and can be controlled by tuning their co-existence. This provides opportunity for materials design and morphological optimization to achieve highest capacitance, fastest switching and highest conductivity.

## 4. Materials and methods

**Materials:** Regioregular P3HT (M1011, Mw = 60,150 g $mol^{-1}$, 97.6% RR) and glass substrates coated with indium tin oxide (ITO) were purchased from Ossila. Regiorandom RRa-P3HT, chloroform, 1,2-dichlorobenzene (anhydrous) and potassium hexafluorophosphate ($KPF_6$, 99.5%) were purchased from Sigma-Aldrich. An Ag/AgCl pellet electrode (Harvard Apparatus, 1mm diameter) was purchased from Warner Instruments.

**Sample preparation**: Substrates were cleaned in subsequent ultrasonic baths of Hellmanex (1% vol), bi-distilled water, acetone and isopropyl alcohol, followed by an UV-$O_3$ treatment (20 minutes at room temperature). The P3HT solutions in chloroform were prepared at concentrations of 5, 10, 20 and 40 mg $ml^{-1}$ and stirred at room temperature overnight. . The RRa-P3HT solution was also prepared with10 mg $ml^{-1}$ in chloroform and stirred at room temperature overnight. P3HT solutions in 1,2-dichlorobenzene were prepared with 10 mg $ml^{-1}$ and 40 mg $ml^{-1}$ and stirred at 80°C overnight. All solutions were spin-coated at room temperature on clean substrates (1000 rpm/60 s, THz samples 700 rpm/60s), on glass coated with ITO for the Vis/NIR spectroelectrochemistry and on quartz for the Raman and THz measurements. Beforehand, gold electrodes were evaporated on the Raman and THz substrates (first layer chromium: 10 nm, second layer gold: 60 nm). Precision tip cotton swabs were used to partially remove the film in order to define an area of 12 $mm^2$ on ITO or to define the channel on the gold patterned substrates. Each sample was then placed in a home-built spectroelectrochemical cell containing both the Ag/AgCl electrode and the electrolyte solution. The spectroelectrochemical results shown in Figure 1 and 3 were performed on a thin P3HT film with about 20 nm thickness and the electrolyte used was 0.1 mol $L^{-1}$ $KPF_6$ in bi-distilled water. The measurements on RRa-P3HT (~ 120 nm thickness) shown in Figure 4 were performed using 0.1 mol $L^{-1}$ $KPF_6$ in acetonitrile electrolyte. For the THz measurements in Figure 6 (~ 270 nm film thickness), 0.1 mol $L^{-1}$ $TBAPF_6$ electrolyte in acetonitrile was used.

**Vis/NIR spectroelectrochemistry:** The spectroelectrochemical measurements were performed using a home-built setup that acquires spectra with a time-resolution of 10 ms.[45] The setup comprises a halogen light source (HL-2000, Ocean Insight), a Flame UV-Vis and a Flame NIR spectrometer (Ocean Insight) triggered by a digital delay/pulse generator (DG535, Stanford Research Systems), and customized

Labview code. The voltage between the Ag/AgCl electrode and the film was applied and measured with a data acquisition card (USB-6008, National Instruments, time resolution of 10 ms). The voltages were applied to the Ag/AgCl quasi-reference, and either the P3HT-coated ITO substrate or the gold electrodes on the OECT-like samples were grounded. The transient response current was converted to voltage using a low noise current preamplifier (SR570, Stanford Research Systems) and measured with the same data acquisition card (USB 6008).

To increase reproducibility, the P3HT films were cycled four times before the experiments. Each pre-cycle consisted of the application of a -0.8 V pulse for 100 s to dope the polymer followed by the application of a +0.4 V pulse for 100 s to dedope the film. Time-resolved measurements of the transient response of the P3HT films were recorded while applying 100 s square pulses of voltage to subsequently dope and dedope the polymer. The applied voltage for doping ranged between -0.1 V to -0.8 V, with steps of -0.1 V. The dedoping bias was always +0.4 V. For each step of voltage, two square pulses were applied to confirm stability (Supplementary Fig. 6). The time-resolved measurements of the transient response of the RRa-P3HT film were recorded while applying 20 s square pulses of voltage to subsequently dope and dedope the polymer. The dedoping bias was always 0 V. The applied voltage for doping ranged between -0.7 V to -1.4 V, with steps of -0.1 V.

**In-situ electrochemical resonant Raman spectroscopy:** The ERRS measurements were conducted using a LabRAM HR800 confocal Raman microscope. The samples were irradiated at 633 nm and an excitation power of 9 mW was employed to avoid sample degradation. The accumulation time was of 8 s for each spectrum. First, the frequency of the lines was calibrated using a silicon reference. Then, the excitation light was focused perpendicular on the used cell to allow precise focusing of the laser on the polymeric film. The samples were prepared as described in the sample preparation, resulting in films with 49 nm thickness and an area of 15 $mm^2$, whereby 1 mm overlap on each electrode was kept. To increase the reproducibility, the films were cycled four times prior to the experiments. Each pre-cycle consisted in the application of a -0.8 V pulse for 60 s to dope the polymer followed by the application of a +0.4 V pulse for 60 s to dedope the film. The steady-state Raman spectra were measured after applying subsequent -0.1 V voltage steps from +0.2 V to -0.8 V with a dwell time of 100 s. Background emissions were subtracted.

**Electrochemical terahertz spectroscopy**: For the THz data, the steady-state response was collected after applying subsequent -0.1 V steps for 120 s, from -0.5 V to -1.8 V. We extracted the short-range THz conductivity of electrochemically doped P3HT via time-domain THz transmission spectroscopy. Femtosecond light pulses from a Ti-Sapphire amplifier system (Astrella, Coherent) with a time duration of ~35 fs, 800 nm center wavelength and repetition rate of 1 kHz were used. Part of this near-infrared light beam was focused on a non-centrosymmetric (110) ZnTe crystal to generate the THz pulses via second order non-linear optical rectification. First, the THz beam was focused on the sample position by off-axis parabolic mirrors and then refocused on a second ZnTe crystal, which was used for detection, via free space electro-optic sampling. In the detection crystal, the THz beam spatially and temporary overlapped with a weak part of the fundamental, the gate beam. For this technique, a quarter wave plate, a Wollaston prism and a balanced photodiode were used. The THz propagation through the detection crystal induced a change in the crystal refractive index and the material became birefringent (Pockel effect). This rotated the polarization of the gate beam as it co-propagated with the THz beam through the crystal. The induced ellipticity in the optical pulses was probed by separating the two vertical polarization components with the quarter waveplate and the Wollaston prism, and their difference was measured in the balanced photodetector. The change in gate beam polarization was linearly proportional to the amplitude of the THz electric field. To measure the THz electric field as a function of time, we changed the time delay between gate and THz pulses via a motorized micrometer translation stage. To increase our set-up sensitivity, we used a recent development of the detection scheme, where the polarization of the gate field was biased, allowing more than one order of magnitude increased sensitivity compared with the conventional electro-optic sampling detection scheme.[55] The THz pulses had a duration of about 1 ps and bandwidth between 0.1 and 2.4 THz, where the latter was limited by the phase mismatch in the non-linear crystals. The THz path was continuously purged with dry nitrogen to eliminate moisture that is strongly absorbed by the THz beam. We were able to extract of the complex optical properties in the ground state. Details about the analysis are described in our previous study.[23]

**MCR data analysis:** Multivariate Curve Resolution (MCR) analysis was performed on the unprocessed data using *pyMCR*, an open-source software library written in Python.[53] The details of the method are described elsewhere.[53] Briefly, the series of measurements is included in a matrix D, and the goal of the MCR is to solve:

$$D = CS^T + \varepsilon \quad (1)$$

where C is the matrix of the species concentrations, S contains their spectral signatures and $\varepsilon$ expresses the error, noise or other features of the measurement series that are not taken into account by the fit. The algorithm uses an iterative alternating regression to minimize $\varepsilon$, and to find the best solution for Equation 1. We first used MCR to find the spectral signatures of each species using the data in Figure 1 from the steady-state Vis/NIR spectroelectrochemical measurements and scaled them according to the absorbance cross sections. The same species were then fixed and used throughout the manuscript to determine the concentrations, including for the time-resolved measurements. Subsequently, the dynamics of the species concentrations were fit by a system of ordinary differential equations using the open-source library *symfit*, written in Python (kinetic modelling).[46] The initial guesses for the rate constant were always random values.

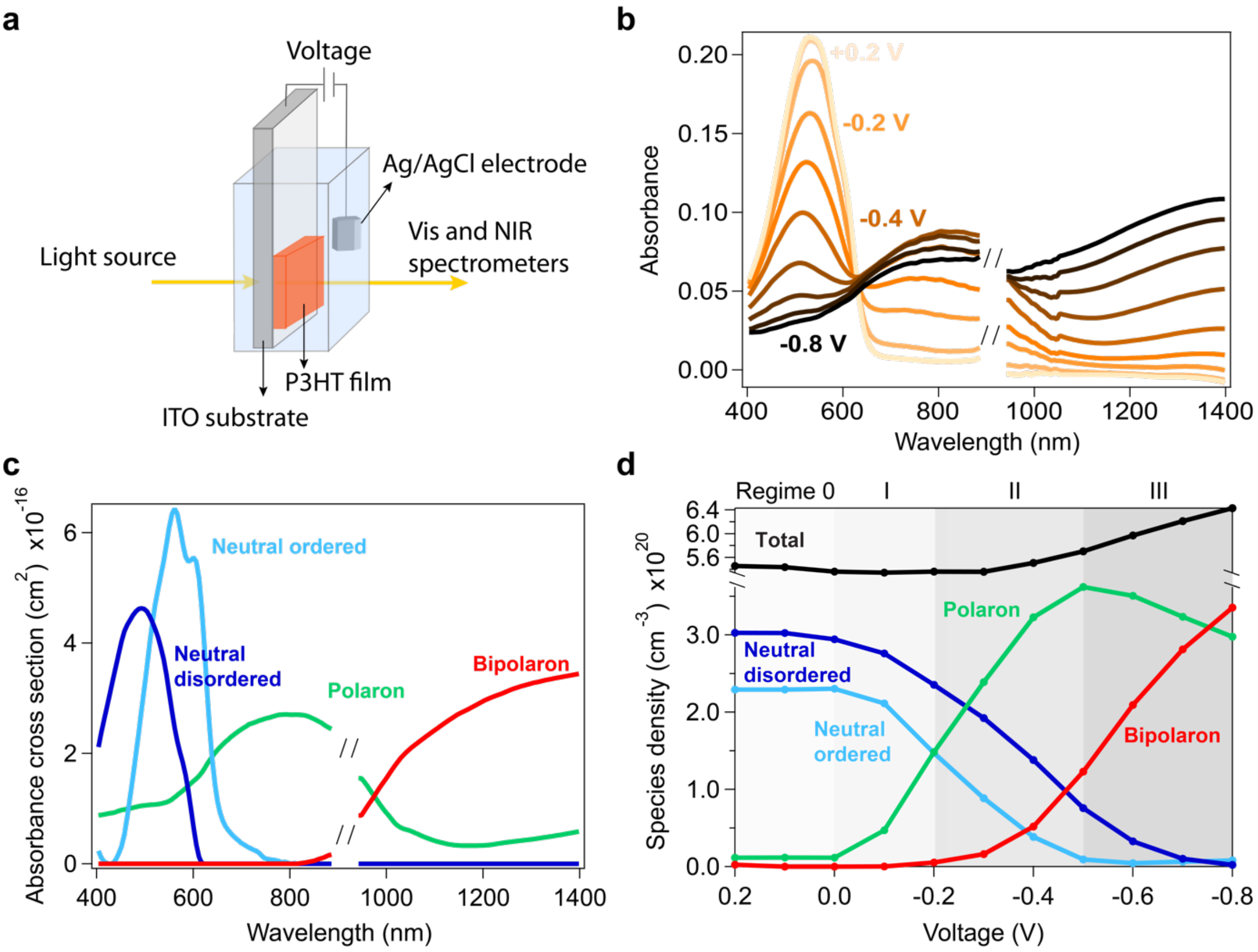

**Figure 1:** a) Experimental setup for the *in-situ* Vis/NIR spectroelectrochemical measurements. b) Steady-state absorbance spectra of P3HT film (cast from chloroform) in aqueous $KPF_6$ electrolyte (0.1 mol $L^{-1}$) upon application of doping voltages ranging from +0.2 V to -0.8 V ($\Delta V$ = -0.1 V, versus Ag/AgCl). The region around 900 nm could not be measured (change between Vis and NIR detector). c) Spectral signature and absorbance cross section of each species obtained from the MCR analysis and d) their corresponding densities as a function of the doping voltage.

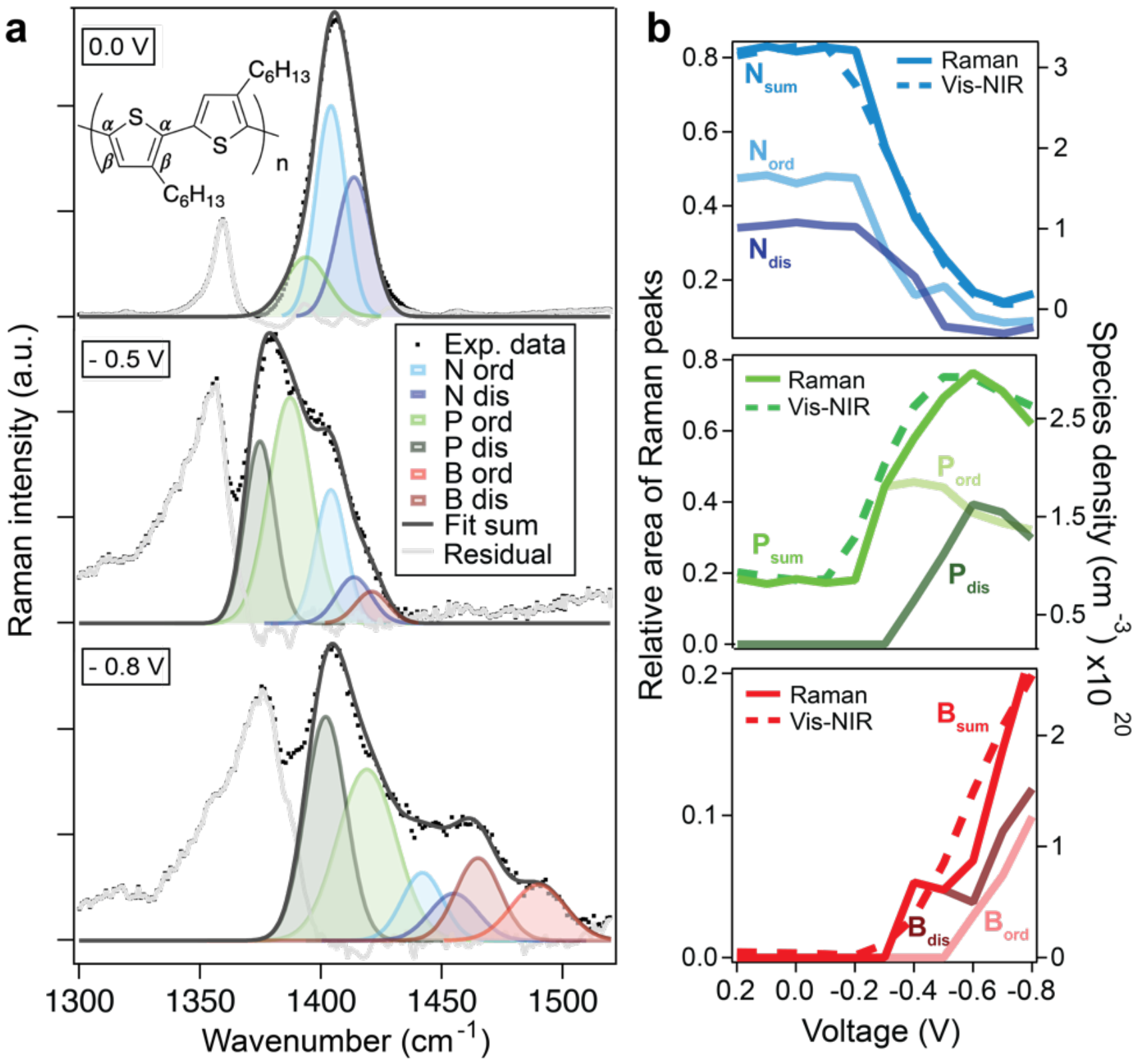

**Figure 2:** a) *In-situ* electrochemical Raman spectra of P3HT film (cast from chloroform) in aqueous $KPF_6$ electrolyte (0.1 mol $L^{-1}$) measured in an 'OECT-like' device with excitation at 633 nm. Experimental spectra (dotted lines) and Gaussian deconvolutions (colored curves) are depicted for voltages applied within regime I (0 V), regime II (-0.5 V) and regime III (-0.8 V). b) Evolution of the relative peak areas of the Raman bands of the different redox species in ordered and disordered domains upon doping from +0.2 V to -0.8 V ($\Delta V$ = -0.1 V, versus Ag/AgCl). The evolution of the total neutral, polaron and bipolaron population from Vis/NIR spectroelectrochemistry on the same device is shown for comparison.

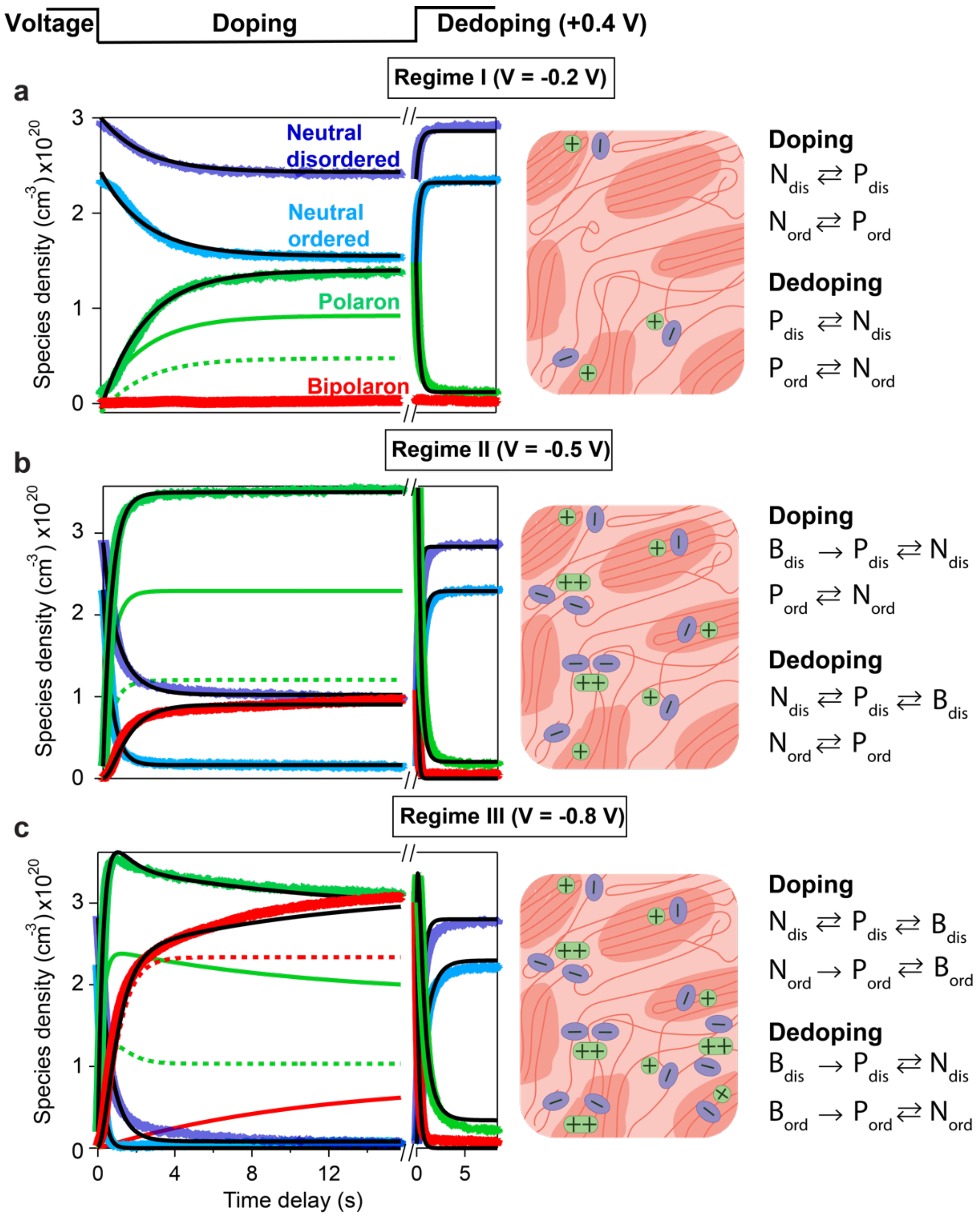

**Figure 3:** Time-resolved species concentrations obtained from VIS/NIR spectroelectrochemistry of a 20 nm P3HT film (cast from chloroform) in aqueous $KPF_6$ electrolyte (0.1 mol $L^{-1}$) upon application of a doping voltage of a) -0.2 V, b) -0.5 V and c) -0.8 V (left) and subsequent dedoping to +0.4 V (right). The dynamics are analysed by kinetic modelling according to the indicated reaction steps (fits shown in black). The simulated ordered (solid line) and disordered (dashed line) subpopulations of polarons (green) and bipolarons (red) are included. For each voltage, a schematic representing the doping in the different morphological domains is depicted (anions in purple and polarons (+) and bipolarons (++) in green).

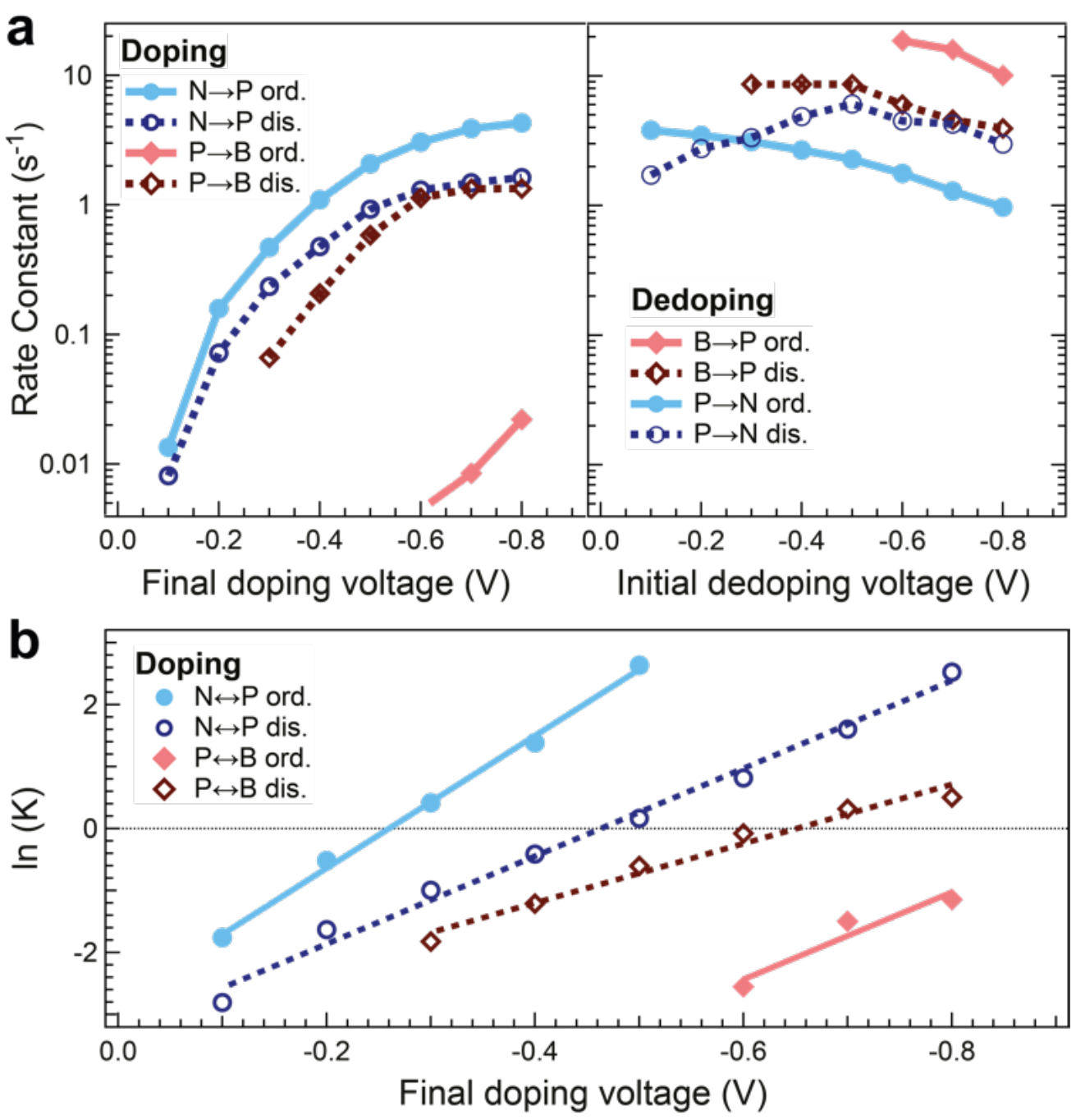

**Figure 4:** a) Evolution of the rate constants obtained by kinetic modelling of the species dynamics from the VIS/NIR spectroelectrochemistry data of P3HT film (cast from chloroform) in aqueous $KPF_6$ electrolyte (0.1 mol $L^{-1}$). The left side shows the doping processes in the ordered and disordered regions of P3HT as a function of oxidation voltage, while the right side shows the corresponding dedoping processes as a function of the initial oxidation voltage (before dedoping to +0.4 V). b) Dependence of the doping equilibrium constants (from the kinetic modelling) on the applied oxidation voltage.

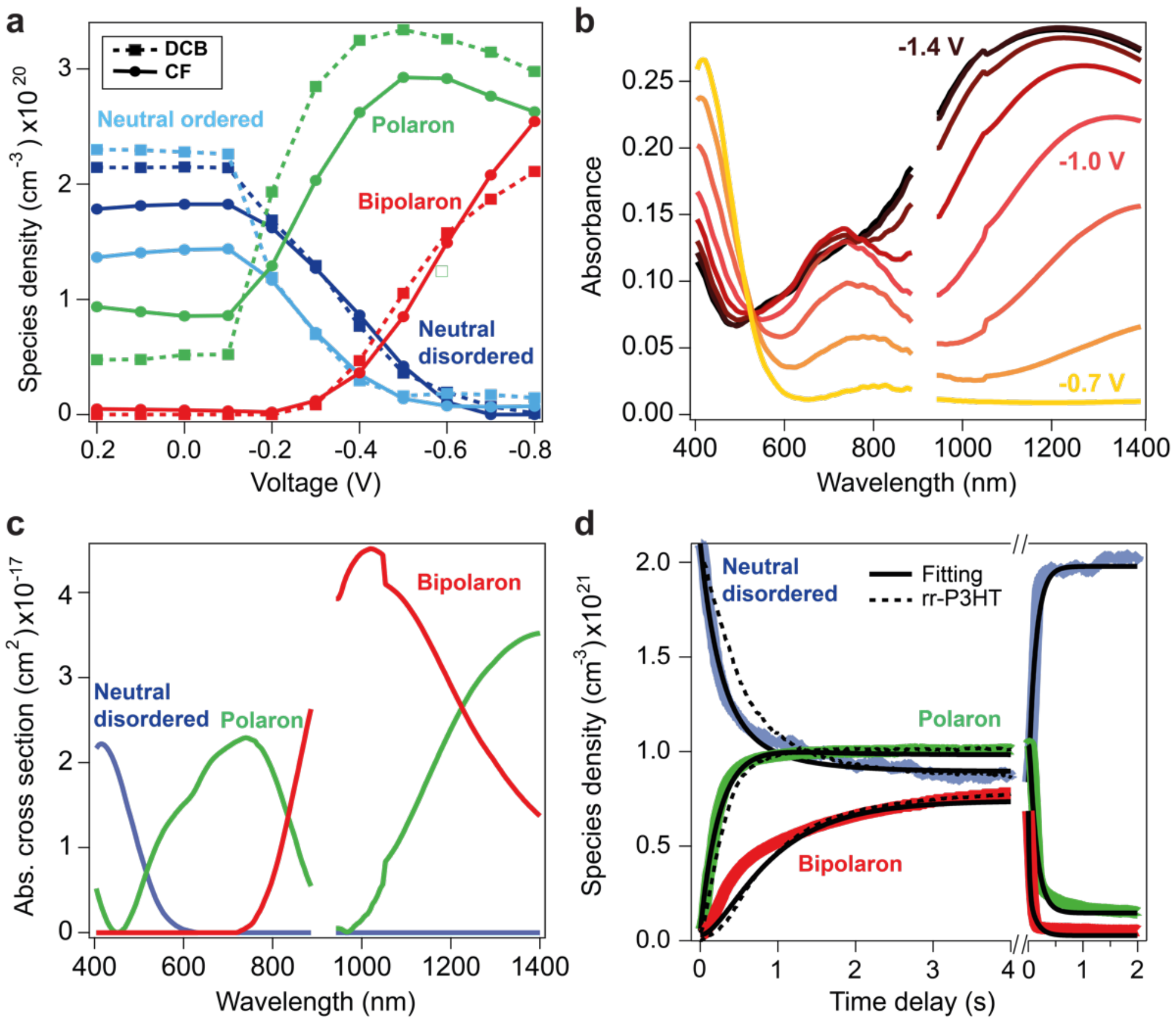

**Figure 5**: a) MCR species densities as a function of doping voltage for regioregular P3HT films cast from 1,2-dichlorobenzene (DCB, squares with dashed lines) and chloroform (CF, circles with solid lines), from VIS/NIR spectro-electrochemistry in aqueous $KPF_6$ electrolyte (0.1 mol $L^{-1}$) measured in 'OECT-like' devices. b) Steady-state absorbance spectra of a 120 nm regiorandom RRa-P3HT film cast on ITO substrate upon application of doping voltages ranging from -0.7 V to -1.4 V ($\Delta V$ = -0.1 V versus Ag/AgCl) in $KPF_6$/acetonitrile electrolyte. c) Corresponding spectral signatures of each species obtained from the MCR analysis. d) Time-resolved species concentrations obtained for the RRa-P3HT film upon application of (left) a doping voltage of -1.2 V and (right) a dedoping voltage of 0 V. For comparison, the doping dynamics at -0.6 V (similar overpotential) of a regioregular P3HT film (rr-P3HT) cast from chloroform are shown.

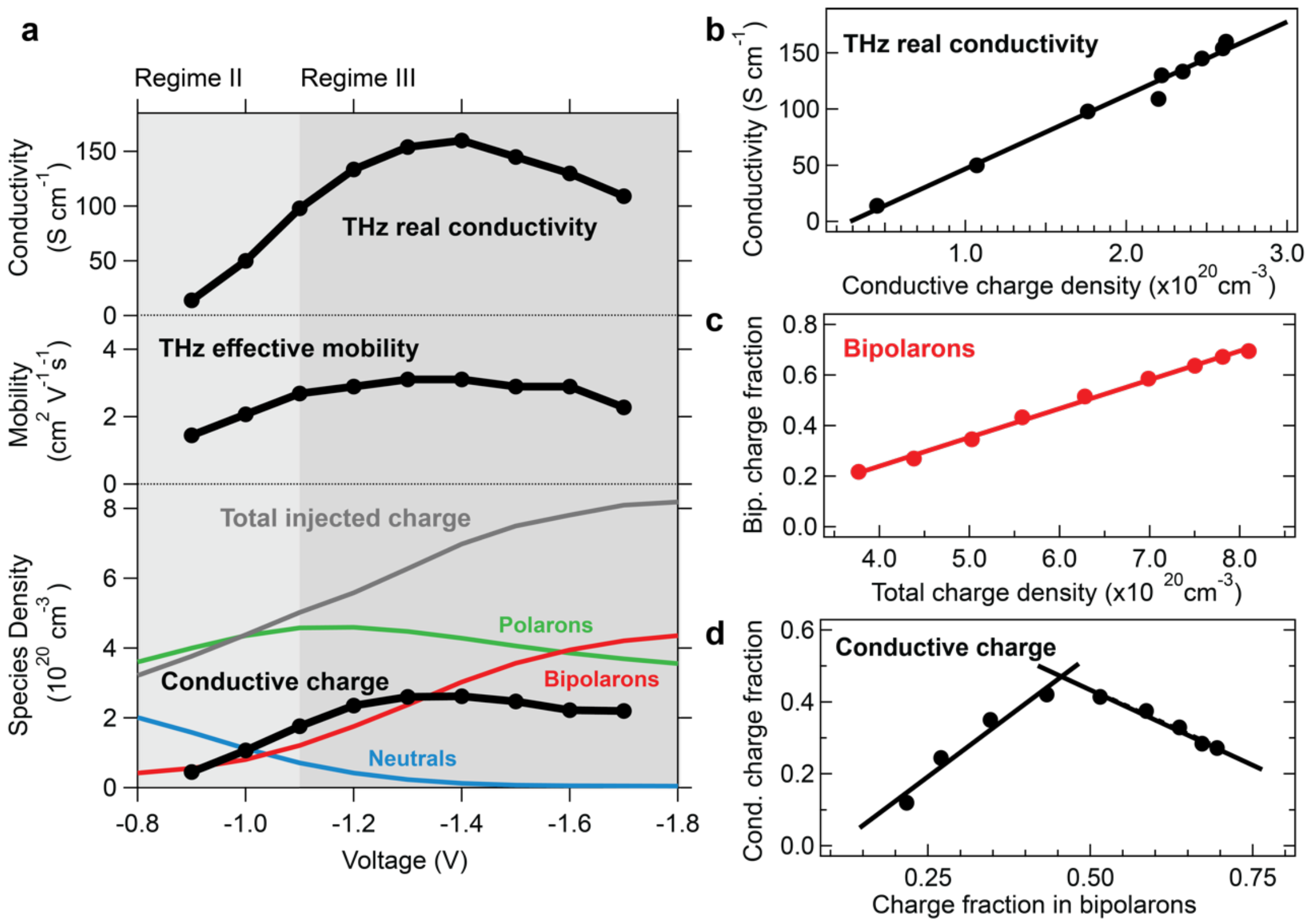

**Figure 6**. a) THz conductivity parameters obtained *in-situ* for a ~270 nm thick P3HT film cast from 1,2-dichlorobenzene in an OECT-like device in $TBAPF_6$/acetonitrile electrolyte. The real conductivity at 1 THz (top), the effective mobility obtained from Drude-Smith analysis (middle), and the density of conductive charges (also from Drude-Smith analysis, bottom) are shown as thick black lines with round markers. In addition, the density of the different redox species (from Vis/NIR spectroelectrochemistry and MCR analysis on the same device) is indicated at the bottom. The total injected charge density is calculated as the polaron plus twice the bipolaron density (P+2xB). b) Real conductivity at 1 THz as a function of conductive charge density. c) Fraction of charge in bipolarons (2B/(2B+P)) as a function of the injected charge density. d) Fraction of conductive charge as a function of the fraction of charge in bipolarons. Solid lines in b)-d) are guides to the eye.

**Table 1:** Rate constants (k in $s^{-1}$) and equilibrium constants (K, unitless) for the electrochemical doping of P3HT at different voltages from -0.1 V to - 0.8V, obtained via kinetic modelling. The equilibrium constants are defined as the forward rate constant divided by the backward rate constant.

| Voltage (V): | - 0.1 | - 0.2 | - 0.3 | - 0.4 | - 0.5 | - 0.6 | - 0.7 | - 0.8 |
|---|---|---|---|---|---|---|---|---|
| $k_{N,ord\rightarrow P,ord}$ | 0.01 | 0.16 | 0.47 | 1.09 | 2.07 | 3.05 | 3.88 | 4.29 |
| $K_{N,ord\rightleftharpoons P,ord}$ | 0.17 | 0.60 | 1.52 | 3.99 | 13.94 | ∞ | ∞ | ∞ |
| $k_{P,ord\rightarrow B,ord}$ | | | | | | 0.004 | 0.01 | 0.02 |
| $K_{P,ord\rightleftharpoons B,ord}$ | | | | | | 0.08 | 0.22 | 0.32 |
| $k_{N,dis\rightarrow P,dis}$ | 0.01 | 0.07 | 0.23 | 0.48 | 0.93 | 1.30 | 1.48 | 1.62 |
| $K_{N,dis\rightleftharpoons P,dis}$ | 0.06 | 0.20 | 0.37 | 0.66 | 1.18 | 2.27 | 4.97 | 12.50 |
| $k_{P,dis\rightarrow B,dis}$ | | | 0.07 | 0.21 | 0.58 | 1.13 | 1.33 | 1.34 |
| $K_{P,dis\rightleftharpoons B,dis}$ | | | 0.16 | 0.30 | 0.55 | 0.92 | 1.37 | 1.65 |

**Table 2:** Rate constants (k in $s^{-1}$) and equilibrium constants (K, unitless) for the electrochemical dedoping of P3HT from different doping voltages (-0.1 V to - 0.8V) to a dedoping voltage of +0.4 V, obtained via kinetic modelling. The equilibrium constants are defined as the forward rate constant divided by the backward rate constant.

| *Voltage (V):* | *- 0.1* | *- 0.2* | *- 0.3* | *- 0.4* | *- 0.5* | *- 0.6* | *- 0.7* | *- 0.8* |
|---|---|---|---|---|---|---|---|---|
| $k_{B,ord\rightarrow P,ord}$ | | | | | | 18.66 | 15.88 | 10.08 |
| $k_{P,ord\rightarrow N,ord}$ | 3.82 | 3.48 | 3.12 | 2.68 | 2.27 | 1.77 | 1.29 | 0.97 |
| $K_{P,ord\rightleftharpoons N,ord}$ | 23.21 | 18.97 | 15.65 | 13.07 | 12.05 | 12.05 | 12.05 | 9.07 |
| $k_{B,dis\rightarrow P,dis}$ | | | 8.63 | 8.63 | 8.63 | 6.01 | 4.60 | 3.93 |
| $k_{P,dis\rightarrow N,dis}$ | 1.72 | 2.75 | 3.34 | 4.85 | 6.07 | 4.49 | 4.24 | 2.98 |
| $K_{P,dis\rightleftharpoons N,dis}$ | ∞ | ∞ | ∞ | ∞ | 202.45 | 56.85 | 31.96 | 31.96 |

Supplementary Information

# Electrochemical Doping in Ordered and Disordered Domains of Conjugated Polymers

Priscila Cavassin, Isabelle Holzer, Demetra Tsokkou, Oliver Bardagot, Julien Réhault, Natalie Banerji*

Department of Chemistry, Biochemistry and Pharmaceutical Sciences, University of Bern, Freiestrasse 3, 3012 Bern, Switzerland

E-mail: natalie.banerji@unibe.ch

## Supplementary Note 1: Multivariant curve resolution (MCR) analysis

The spectro-electrochemical data was decomposed into species and concentrations using MCR-ALS analysis. As shown in Fig. S1a, the absorbance of the P3HT film on ITO in the near infrared is reproducibly negative due to reflection effects. Therefore, to obtain a more precise MCR decomposition, the differential absorbance in relation to the undoped film was used. Fig. S1b shows the differential data and MCR fitting results. The concentrations obtained are shown in Fig. S1c. To facilitate the interpretation of the results, we have decomposed the absolute spectrum of the dedoped film, as shown in Fig. S1a, and obtained the initial density of neutral disordered, ordered and polarons. These values were then added to the curves shown in Fig.1c to obtain the final densities shown in Fig. S1d.

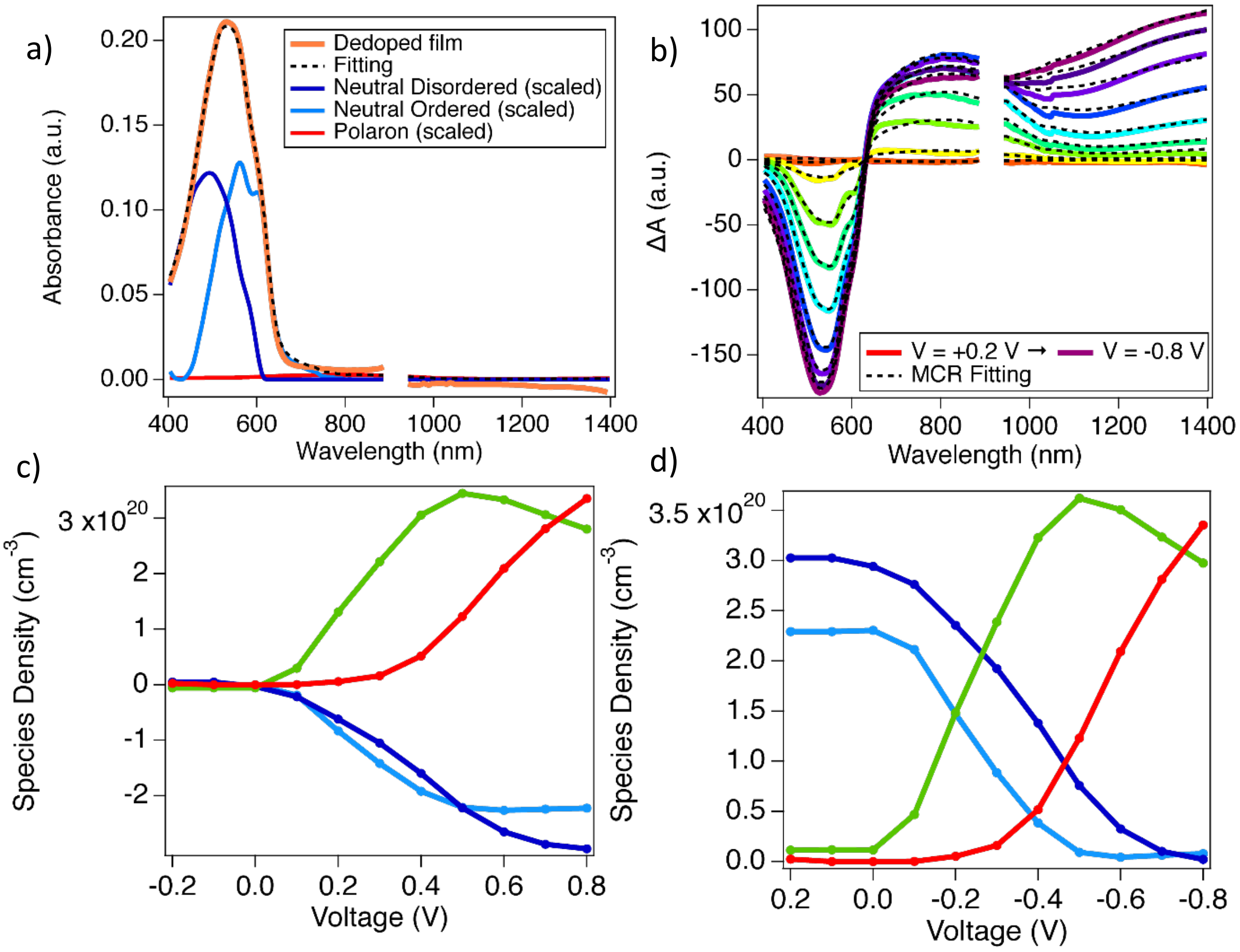

**Supplementary Figure 1**: **Multivariant curve resolution (MCR) analysis.** a) Absolute absorbance of the dedoped P3HT film on ITO (in red) measured in aqueous NaCl at open-circuit voltage after 5 doping and dedoping cycles. The MCR fitting result is shown in dashed line and the decomposed scaled spectral components are also depicted. b) Steady-state in-situ differential absorbance spectra of the P3HT film upon application of doping voltages ranging from +0.2 V to -0.8 V (ΔV = -0.1 V *vs.* Ag/AgCl). The MCR fitting is shown in dashed lines. c) Species density obtained with the MCR decomposition of the differential absorbance, and d) adjusted densities after adding the initial densities of each species obtained with the fitting in a).

## Supplementary Note 2: Absorbance cross section determination

Because the MCR does not provide information on the absolute value of the concentrations, we have used the charge injected into the polymer film (from chronoamperometry) to calculate the absorbance cross section of each species and their absolute density. For each applied voltage pulse, we obtain the injected charge by integrating the measured current, as we have described in detail in a previous study.[1] The absorbance at the maximum signature of the neutral (sum of ordered and disordered as discussed below, Equation S9), the polaron (maximum of the P2 band) and bipolaron species divided by the film thickness is plotted against the injected carrier density in Fig. S2a-c. This was recorded for samples of different film thickness, cast from chloroform or dichlorobenzene, and for the two used architectures (ITO-based or OECT-like). We observed that even though the investigated samples have similar doping levels, as can be seen in the absorbance, they have different injected carrier densities. Specifically, we noticed that thinner samples (less than 100 nm) are more subject to this variation, which is most severe for the bipolarons. This is likely because the measured current is the sum of capacitive, faradaic and leakage contributions. Although we attempt to eliminate the leakage contribution during the analysis,[1] the thinner films have less absolute charge participating in the charge transfer process, so that they are more sensitive to error. For this reason, we choose to obtain the absorbance cross section using the data obtained for three distinct films, with higher thicknesses of 110 nm, 130 nm, and 270 nm.

The relation between absorbance and species density is described by:

$$\frac{A}{t} = \frac{\sigma}{2.3} N \qquad (S1)$$

where A is the species absorbance, t is the film thickness, $\sigma$ is the absorbance cross section and N is the species concentration.[2] The absorption cross section can therefore be deduced from the slope of the graphs in Fig. S2 (see details below). The slopes are similar for the three used film thicknesses (including ITO and OECT architecture and both casting solvents), confirming the robustness of the analysis and allowing to link this parameter during a global liner fitting of the three curves.

For the polarons, we assume that the injected carrier density is equal to the density of polarons for the lower injected charge densities (< $2 \cdot 10^{20}$ $cm^{-3}$), where the concentration of bipolarons is close to zero. Therefore, by linearly fitting these points, we obtain the absorbance cross section of the polaron. Having this value, we calculate the density of polarons also at the higher doping levels for each sample. Considering that all injected carriers form either polarons and bipolarons, and that the bipolarons are formed by two injected charges, we obtain the density of bipolarons using:

$$N_B = \frac{q - N_P}{2} \qquad (S2)$$

where $N_B$ and $N_P$ are the density of bipolarons and polarons (determined above), and q is the injected carrier density. The absorbance of the bipolarons for the films is plotted versus the density of bipolarons in Supplementary Fig. 2d, and the absorbance cross section for the bipolarons is found from the slope using Equation S1.

Finally, each injected charge leads to the loss of one neutral species in the regime where only polarons and no bipolarons are formed. In the case of RR-PH3T, where there are two different neutral species that are simultaneously decaying, it is not possible to simply use Equation S1 to obtain the coefficients. We now derive an expression to decouple both contributions. Considering Equation S1, we have the following for neutral disordered ordered species:

$$\frac{A_{N,dis}}{t} = \frac{\sigma_{N,dis}}{2.3} N_{N,dis} \qquad (S3)$$

$$\frac{A_{N,ord}}{t} = \frac{\sigma_{N,ord}}{2.3} N_{N,ord} \qquad (S4)$$

Clark et al. have demonstrated that for RR-P3HT, the absorbance cross section of the neutral in ordered and disordered domains are related by the expression

$$\sigma_{N,ord} = 1.39\sigma_{N,dis} \qquad (S5)$$

Thus, rewriting S4 using relation S5 we obtain

$$\frac{A_{N,ord}}{1.39\, t} = \frac{\sigma_{N,dis}}{2.3} N_{N,ord} \qquad (S6)$$

Summing expressions S3 and S6:

$$\frac{A_{N,dis}}{t} + \frac{A_{N,ord}}{1.39\, t} = \frac{\sigma_{N,dis}}{2.3} \left(N_{N,dis} + N_{N,ord}\right) \qquad (S7)$$

Considering that before the bipolarons start to be formed one injected carrier converts one neutral into one polaron:

$$N_{N,initial} - q = N_{N,dis} + N_{N,ord} \qquad (S8)$$

where $N_{N,initial}$ is the initial amount of neutrals. Thus,

$$\frac{A_{N,dis}}{t} + \frac{A_{N,ord}}{1.39\, t} = -\frac{\sigma_{N,dis}}{2.3} q + \frac{\sigma_{N,dis}}{2.3} N_{N,initial} \qquad (S9)$$

Based on this relation, the sum of the neutral absorbance is found according to the left side of Equation S9 and plotted versus the injected carrier density in Fig. S2a. The absorbance cross section of the neutral disordered species is found from the slope at low injected charge densities (before bipolaron formation). Finally, using Equation S5, the neutral ordered absorbance cross section is obtained. For RRa-P3HT, only neutral disordered species are present, so that the coefficient is simply obtained by plotting the neutral absorbance versus the injected charge and linearly fitting the data (for a 125 nm film cast from chloroform).

All absorbance cross sections are summarized in Table S1. Once the cross-sections were determined, the intensity of the spectral components obtained by MCR was scaled to those coefficients. By fixing the adjusted species and running the MCR again, the absolute densities of the species were obtained.

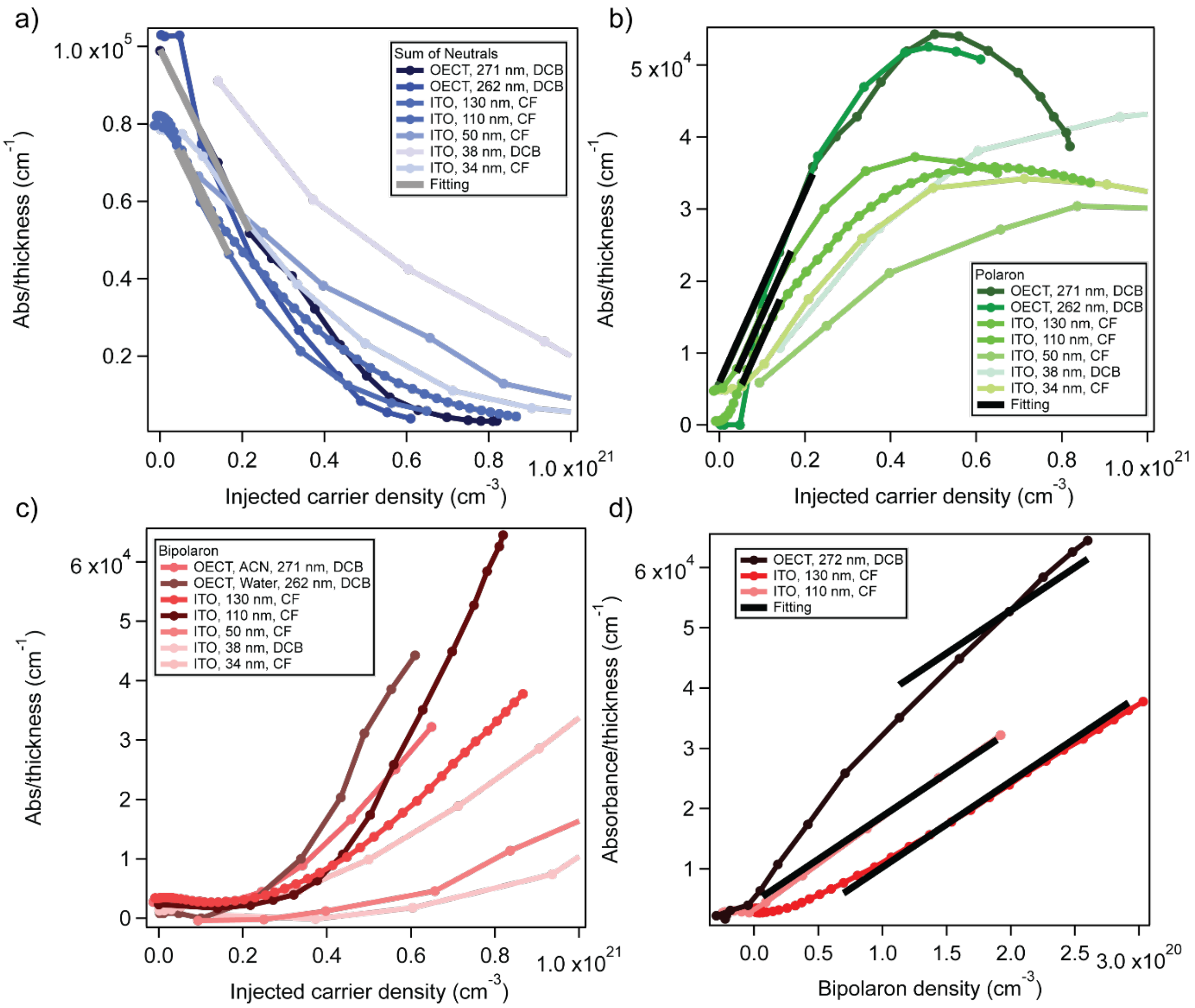

**Supplementary Figure 2: Absorbance cross section determination.** a) Sum of neutral, b) polaron and c) bipolaron absorbance divided by film thickness versus injected carrier density for several films, spin coated from solutions in chloroform (CF) or 1,2-dichlorobenzene (DCB). It is indicated if the films were deposited on ITO or on an OECT-like device structure. The global fittings for the thicker films of 110 nm, 130 nm and 270 nm (used for the extinction coefficient analysis) are indicated by thicker solid lines. d) Absorbance divided by thickness versus the density of bipolarons.

**Supplementary Table 1: Coefficients obtained for the RR and RRa-P3HT films**

| | Absorbance cross section ($cm^2$) | |
|---|---|---|
| | RR-P3HT | RRa-P3HT |
| Neutral disordered | $6.76 \cdot 10^{-16}$ | $2.02 \cdot 10^{-17}$ |
| Neutral Ordered | $4.87 \cdot 10^{-16}$ | - |
| Polaron | $2.97 \cdot 10^{-16}$ | $4.95 \cdot 10^{-17}$ |
| Bipolaron | $3.27 \cdot 10^{-16}$ | $3.48 \cdot 10^{-17}$ |

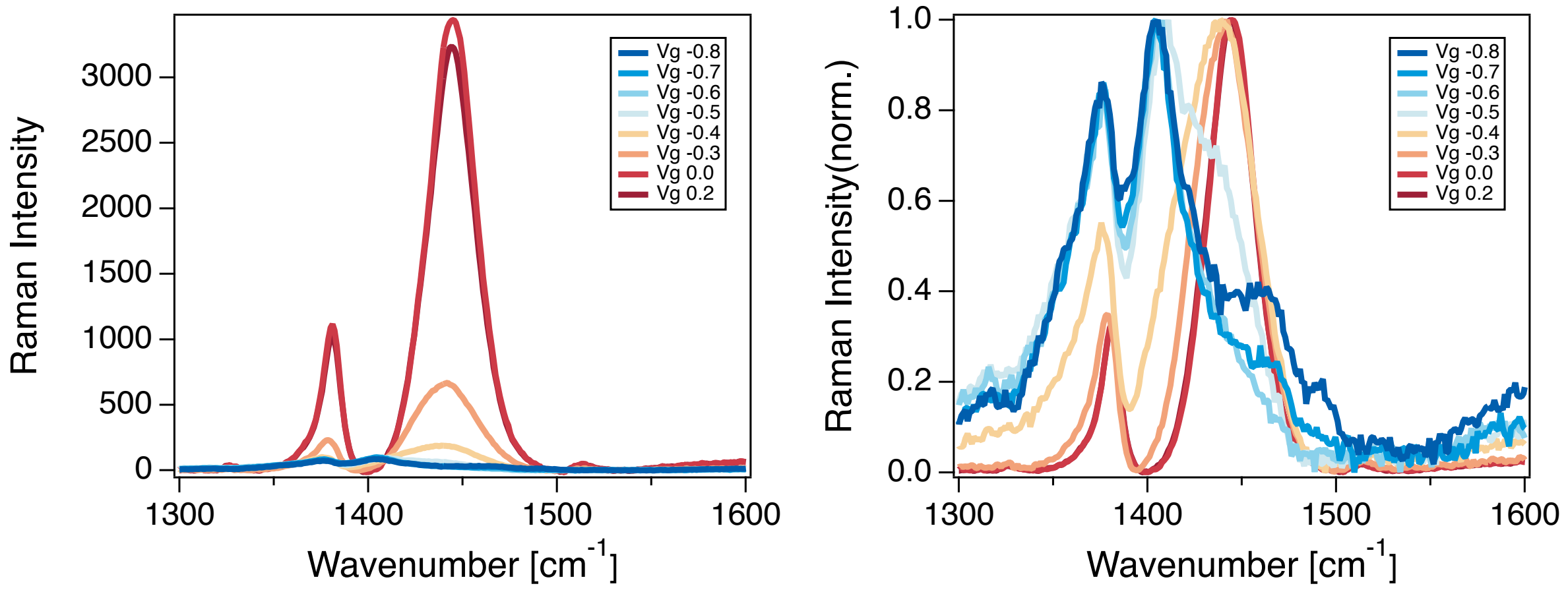

**Supplementary Figure 3**: **Non-normalized vs normalized Raman spectra.** Comparison of the normalized electrochemical resonant Raman spectra and the non-normalized ones, with the background subtracted from both. The normalized spectra were additionally normalized to the $C_\alpha = C_\beta$ peak.

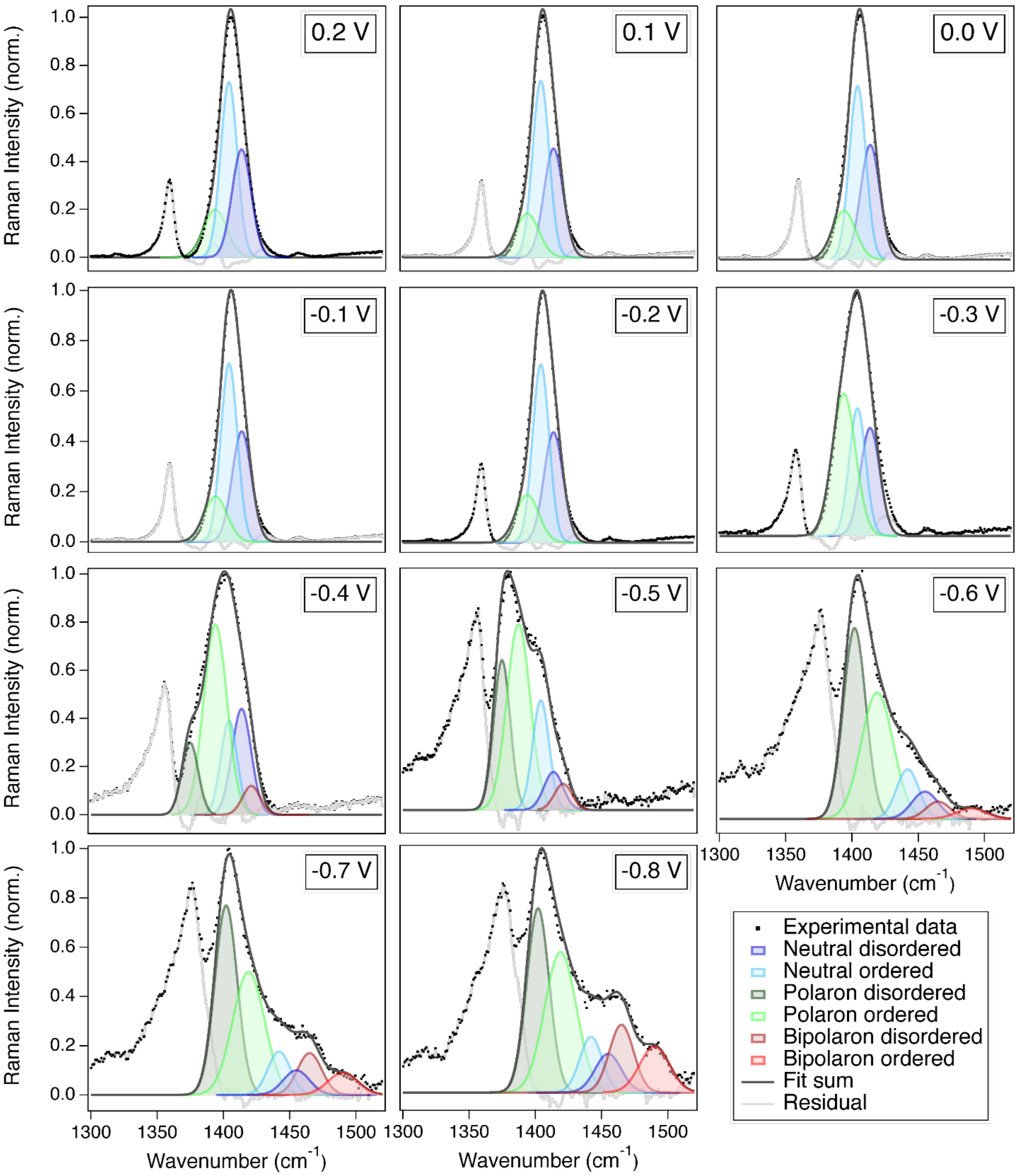

**Supplementary Figure 4: Raman spectra deconvolution.** Gaussian fitting to the Raman spectra for all voltages (from 0.2 to -0.8 V) to determine the different species present.

**Supplementary Table 2: Raman peaks fitting results obtained for all the voltages**
The spectra were fit with Gaussian bands representing the neutral, polaron and bipolaron species in the ordered and disordered phases of the film (see main text). Fitting parameters: The individual peak positions were constant for all voltages except for the ordered polarons, the peak widths were constant for all the modes throughout the different doping states, the peak heights were unconstrained to represent the evolution of the species and order as a function of voltage (but are subject to the normalization of the Raman bands).

| Voltage *(V)* | Neutral ordered | Neutral disordered | Polaron ordered | Polaron disordered | Bipolaron disordered | Bipolaron ordered |
|---|---|---|---|---|---|---|
| | Position, Width, Height | | | | | |
| 0.2 | 1455, 14, 0.45 | 1442, 12, 0.73 | 1428, 17, 0.20 | 0, 0, 0 | 0, 0, 0 | 0, 0, 0 |
| 0.1 | 1455, 14, 0.45 | 1442, 12, 0.73 | 1428, 17, 0.18 | 0, 0, 0 | 0, 0, 0 | 0, 0, 0 |
| 0.0 | 1455, 14, 0.47 | 1442, 12, 0.71 | 1428, 17, 0.20 | 0, 0, 0 | 0, 0, 0 | 0, 0, 0 |
| -0.1 | 1455, 14, 0.44 | 1442, 12, 0.71 | 1428, 17, 0.18 | 0, 0, 0 | 0, 0, 0 | 0, 0, 0 |
| -0.2 | 1455, 14, 0.44 | 1442, 12, 0.71 | 1428, 17, 0.19 | 0, 0, 0 | 0, 0, 0 | 0, 0, 0 |
| -0.3 | 1455, 14, 0.44 | 1442, 12, 0.52 | 1428, 17, 0.58 | 0, 0, 0 | 0, 0, 0 | 0, 0, 0 |
| -0.4 | 1455, 14, 0.39 | 1442, 12, 0.44 | 1428, 17, 0.79 | 1402, 12, 0.30 | 1465, 13, 0.12 | 0, 0, 0 |
| -0.5 | 1455, 14, 0.46 | 1442, 12, 0.16 | 1419, 17, 0.78 | 1402, 12, 0.63 | 1465, 13, 0.11 | 0, 0, 0 |
| -0.6 | 1455, 14, 0.20 | 1442, 12, 0.11 | 1419, 17, 0.51 | 1402, 12, 0.77 | 1465, 13, 0.07 | 1490, 16, 0.04 |
| -0.7 | 1455, 14, 0.17 | 1442, 12, 0.10 | 1419, 17, 0.50 | 1402, 12, 0.77 | 1465, 13, 0.17 | 1490, 16, 0.09 |
| -0.8 | 1455, 14, 0.23 | 1442, 12, 0.16 | 1419, 17, 0.58 | 1402, 12, 0.76 | 1465, 13, 0.28 | 1490, 16, 0.19 |

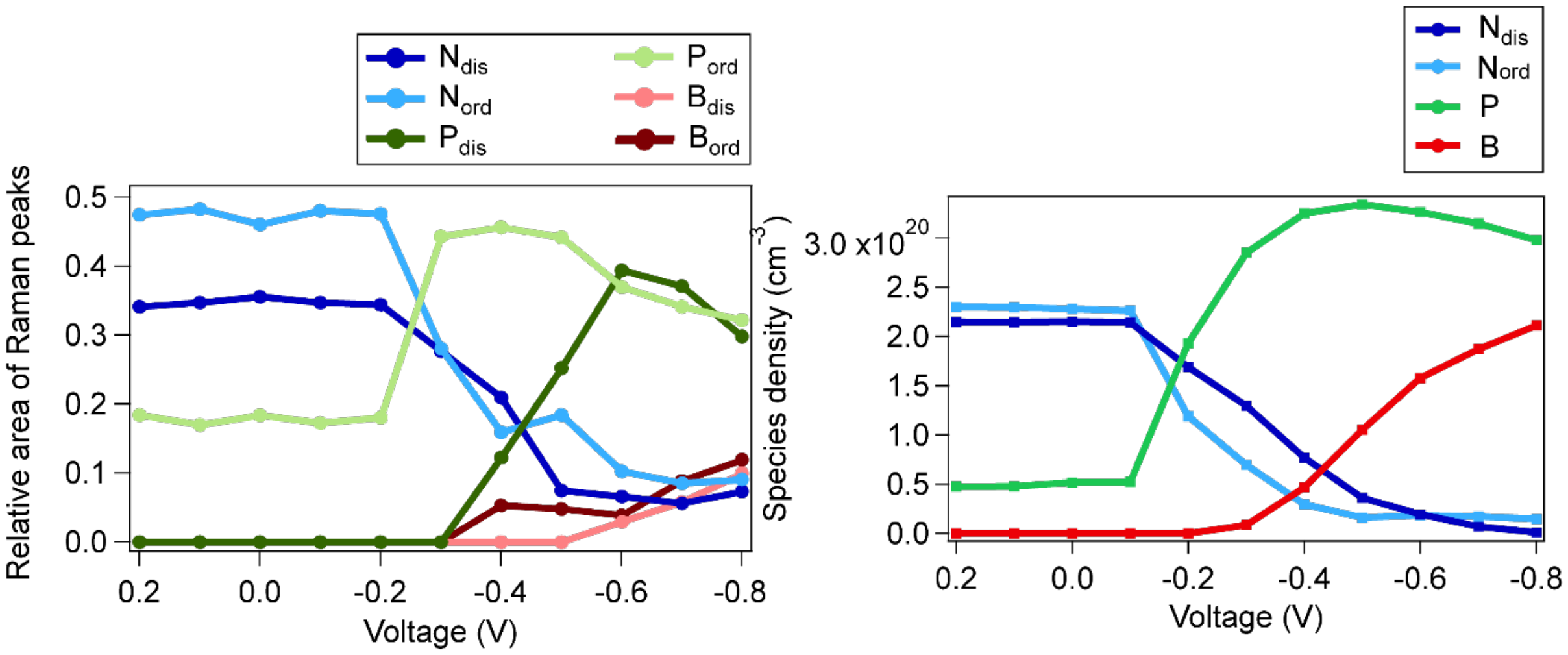

**Supplementary Figure 5**: **Raman and MCR species density comparison.** Comparison between the relative area of the Gaussian Raman bands (normalized by the total area of all bands) and the concentrations of the corresponding species obtained via MCR of the Vis-NIR absorbance spectra. Both experiments were done on the same film in OECT-like configuration, under the same conditions. In the Vis-NIR data, we cannot distinguish between ordered and disordered charged species.

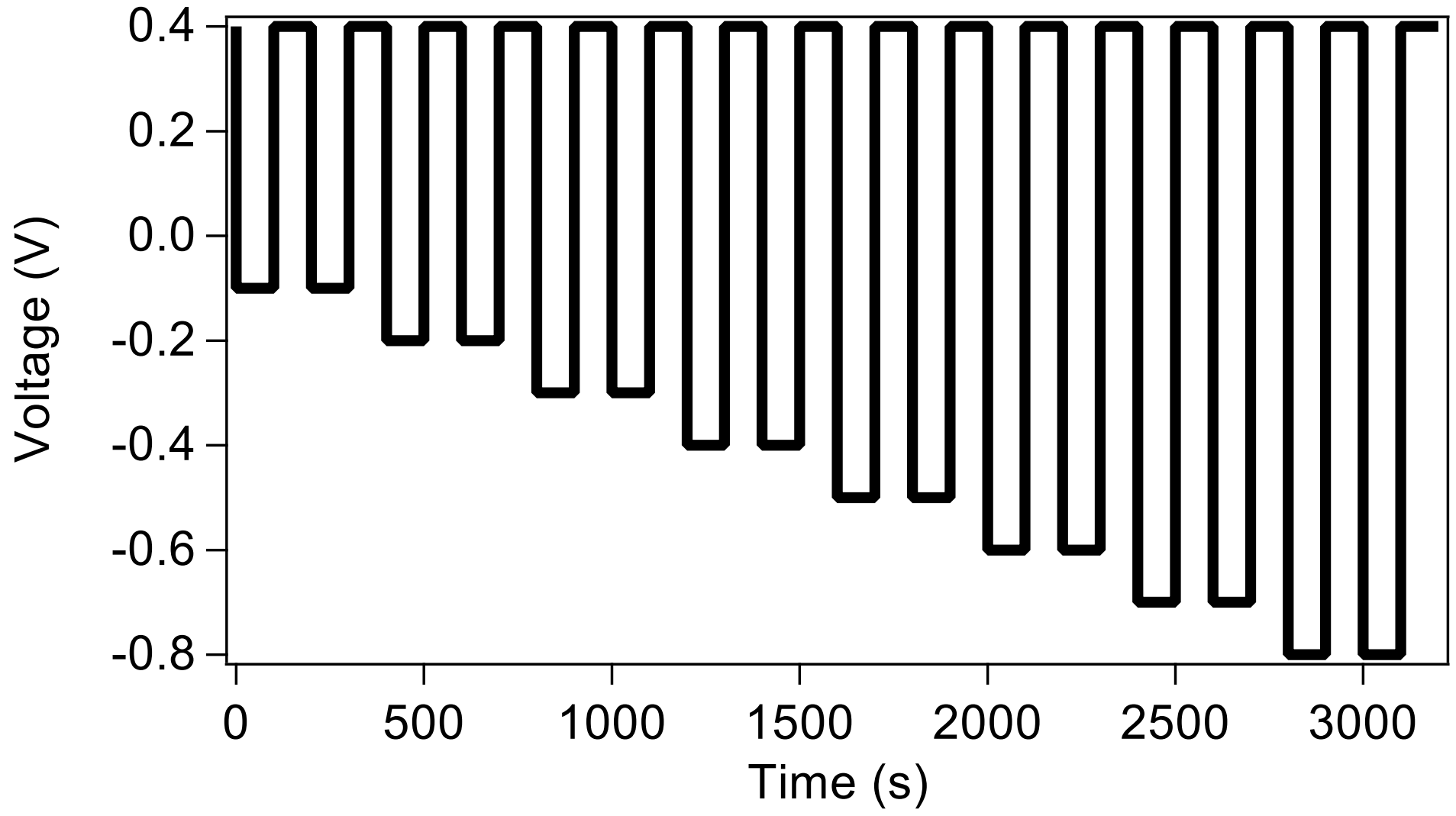

**Supplementary Figure 6**: **Voltage profile of the time-resolved spectroelectrochemical measurements.** The graph shows the voltage applied to the film during the spectroelectrochemical measurements. Subsequent square voltage pulses of 100s were applied to the film.

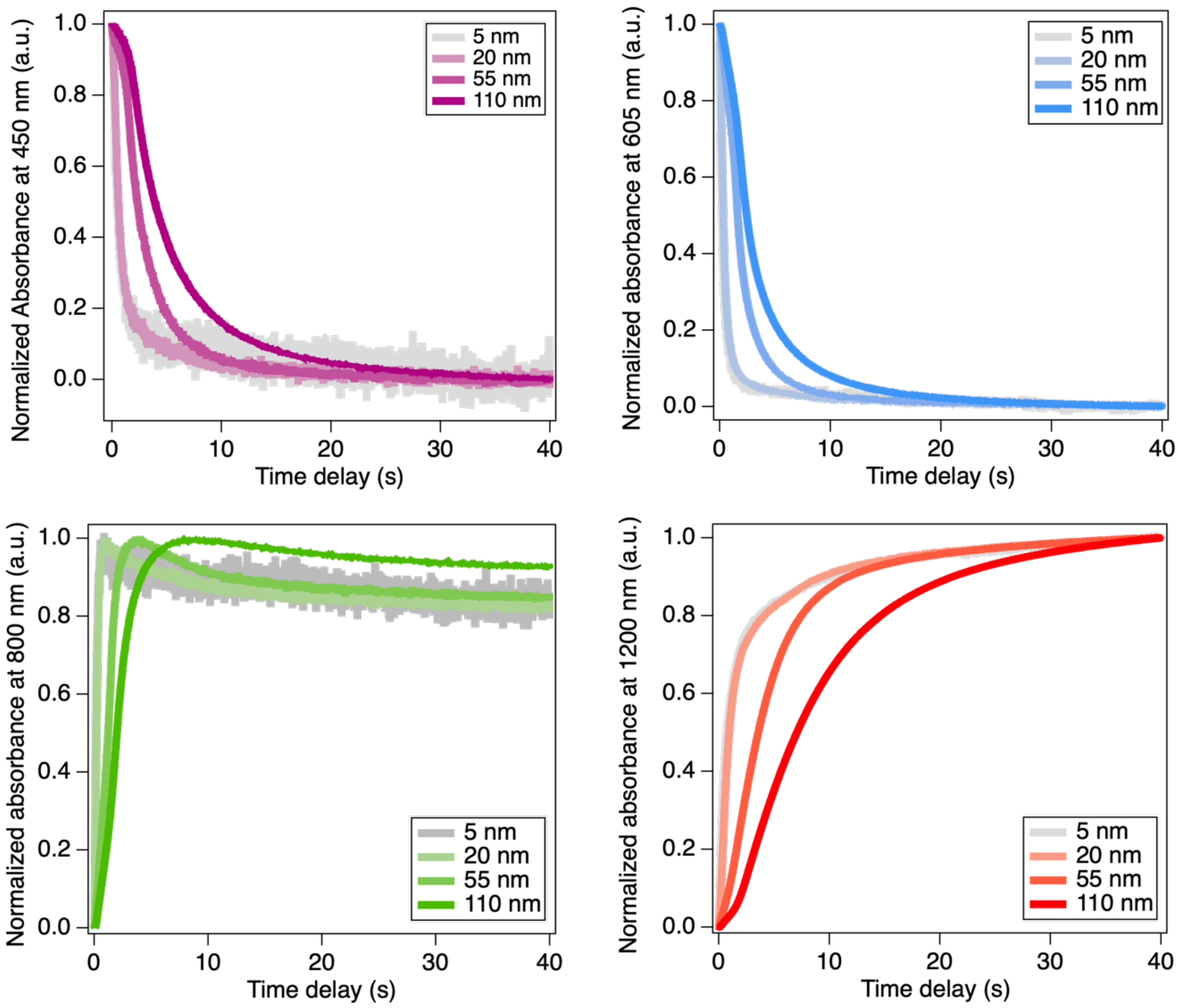

**Supplementary Figure 7**: **Changes in the dynamics for different film thicknesses.** Comparison of the VIS-NIR spectroelectrochemical doping dynamics (at -0.8 V) for P3HT films with 5, 20, 55 and 110 nm thickness. Here, a simple wavelength sampling was used. The purple colour is at the maximum of the neutral disordered band, the blue of the neutral ordered band and the green and red of the polaron and bipolaron bands, respectively. The two thinner films have similar dynamics, which is an indication that in this regime the ionic diffusion is not a limiting factor.

**Supplementary Note 3: Exponential fitting of the doping dynamics**

The time-resolved spectroelectrochemical doping dynamics are shown in Figure S7 for the different redox species at all applied potentials. We find that the curves can generally be analysed with a bi-exponential fit and offset according to the equations shown below (D= disordered neutral, O = ordered neutral, P = polaron, B = bipolaron). Depending on the regime, some parameters were held at zero and similar time constants were linked during the fit. The fits are shown in Figure S7 and S8 (log scale) and the parameters summarized in Table S3.

$$D(t) = D_1 exp\left(-t/_{\tau_{D1}}\right) + D_2 exp\left(-t/_{\tau_{D2}}\right) + y_D \qquad (S10)$$
$$O(t) = O_1 exp\left(-t/_{\tau_{O1}}\right) + O_2 exp\left(-t/_{\tau_{O2}}\right) + y_O \qquad (S11)$$
$$P(t) = P_1 exp\left(-t/_{\tau_{P1}}\right) + P_2 exp\left(-t/_{\tau_{P2}}\right) + y_P \qquad (S12)$$
$$B(t) = B_1 exp\left(-t/_{\tau_{B1}}\right) + B_2 exp\left(-t/_{\tau_{B2}}\right) + y_B. \qquad (S13)$$

**Supplementary Table 3**: Parameters obtained by fitting the doping dynamics. Amplitudes are expressed as a % of the total amplitude.

| ***V (V)*** | ***-0.1*** | ***-0.2*** | ***-0.3*** | ***-0.4*** | ***-0.5*** | ***-0.6*** | ***-0.7*** | ***-0.8*** |
|---|---|---|---|---|---|---|---|---|
| $\tau_{D1}(s)$ | **5.5** (7%) | **2.0** (22%) | **1.0** (33%) | **0.7** (48%) | **0.6** (63%) | **0.5** (77%) | **0.5** (84%) | **0.5** (87%) |
| $\tau_{D2}(s)$ | | | **13.2** (4%) | **8.0** (5%) | **9.2** (7%) | **7.6** (8%) | **4.6** (9%) | **4.1** (11%) |
| $y_D$ $(10^{20} cm^{-3})$ | **2.8** (93%) | **2.4** (78%) | **2.0** (63%) | **1.5** (46%) | **0.9** (30%) | **0.5** (15%) | **0.2** (7%) | **0.1** (2%) |
| $\tau_{O1}(s)$ | **5.5** (10%) | **2.0** (42%) | **0.8** (66%) | **0.5** (83%) | **0.4** (93%) | **0.3** (98%) | **0.2** (99%) | **0.2** (98%) |
| $\tau_{O2}(s)$ | | | **5.5** (3%) | **5.9** (2%) | **4.3** (2%) | | | |
| $y_O$ $(10^{20} cm^{-3})$ | **2.1** (90%) | **1.5** (58%) | **1.0** (31%) | **0.5** (15%) | **0.2** (5%) | **0.0** (2%) | **0.0** (1%) | **0.1** (2%) |
| $\tau_{P1}(s)$ | **5.5** (-100%) | **2.0** (-100%) | **0.8** (-92%) | **0.5** (-95%) | **0.4** (-96%) | **0.3** (-100%) | **0.2** (-100%) | **0.2** (-100%) |
| $\tau_{P2}(s)$ | | | **5.5** (-8%) | **5.9** (-5%) | **4.3** (-4%) | **5.6** (1%) | **4.9** (7%) | **4.7** (14%) |
| $y_P$ $(10^{20} cm^{-3})$ | **0.5** (100%) | **1.4** (100%) | **2.3** (100%) | **3.1** (100%) | **3.0** (100%) | **3.6** (99%) | **3.4** (93%) | **3.1** (86%) |
| $\tau_{B1}(s)$ | | | **1.0** (-100%) | **0.7** (-85%) | **0.6** (-84%) | **0.6** (-82%) | **0.5** (-78%) | **0.5** (-70%) |
| $\tau_{B2}(s)$ | | | | **8.0** (-15%) | **9.2** (-16%) | **5.6** (-18%) | **4.9** (-22%) | **4.7** (-30%) |
| $y_B$ $(10^{20} cm^{-3})$ | | | **0.1** (100%) | **0.4** (100%) | **1.0** (100%) | **1.7** (100%) | **2.5** (100%) | **3.1** (100%) |

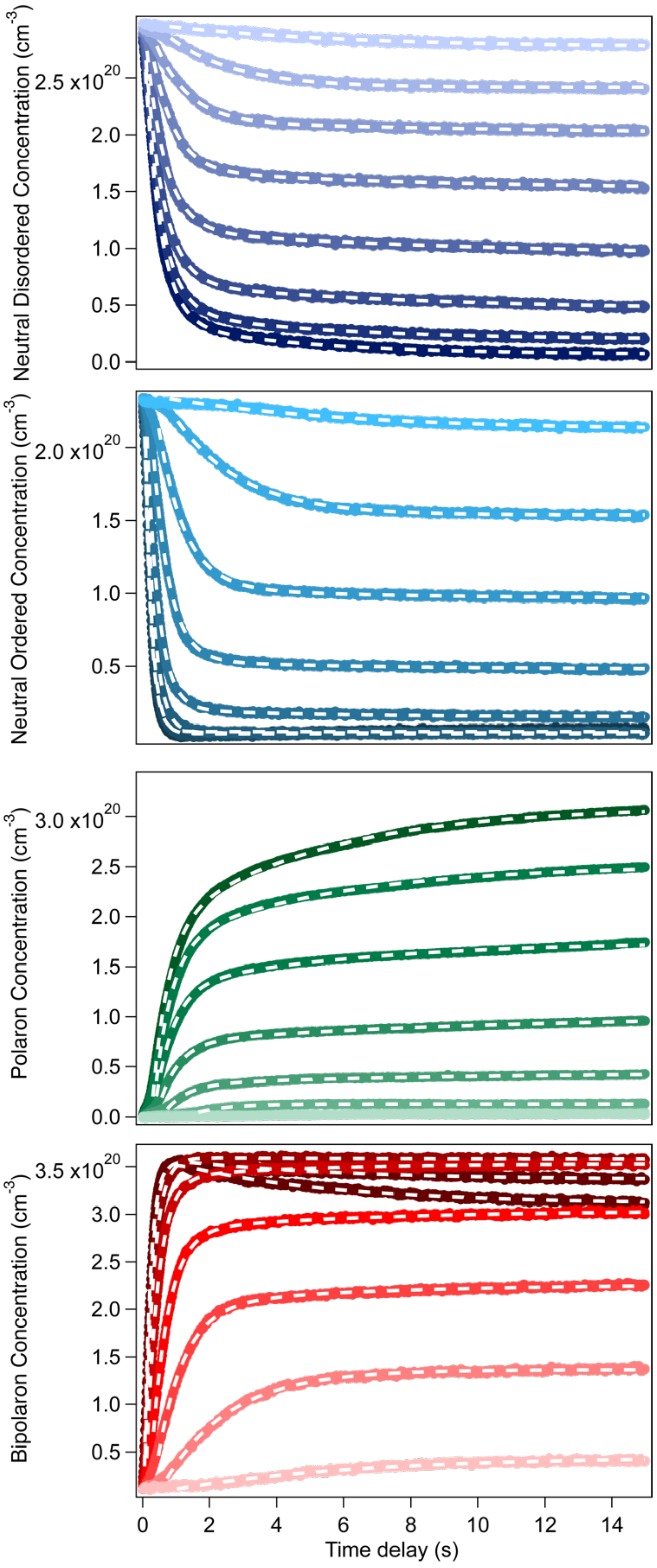

**Supplementary Figure 8**: **Exponential fitting of the time-resolved doping dynamics.** Exponential Time-resolved densities of the neutral disordered (darker blue), neutral ordered (lighter blue), polaron (green) and bipolaron (red) species obtained from the MCR analysis of the spectroelectrochemistry data when switching the voltage according to Figure S6 during doping. From lighter to darker are the different applied voltages ranging between -0.1 V and -0.8 V, with steps of -0.1. The biexponential fittings are shown as dashed white lines.

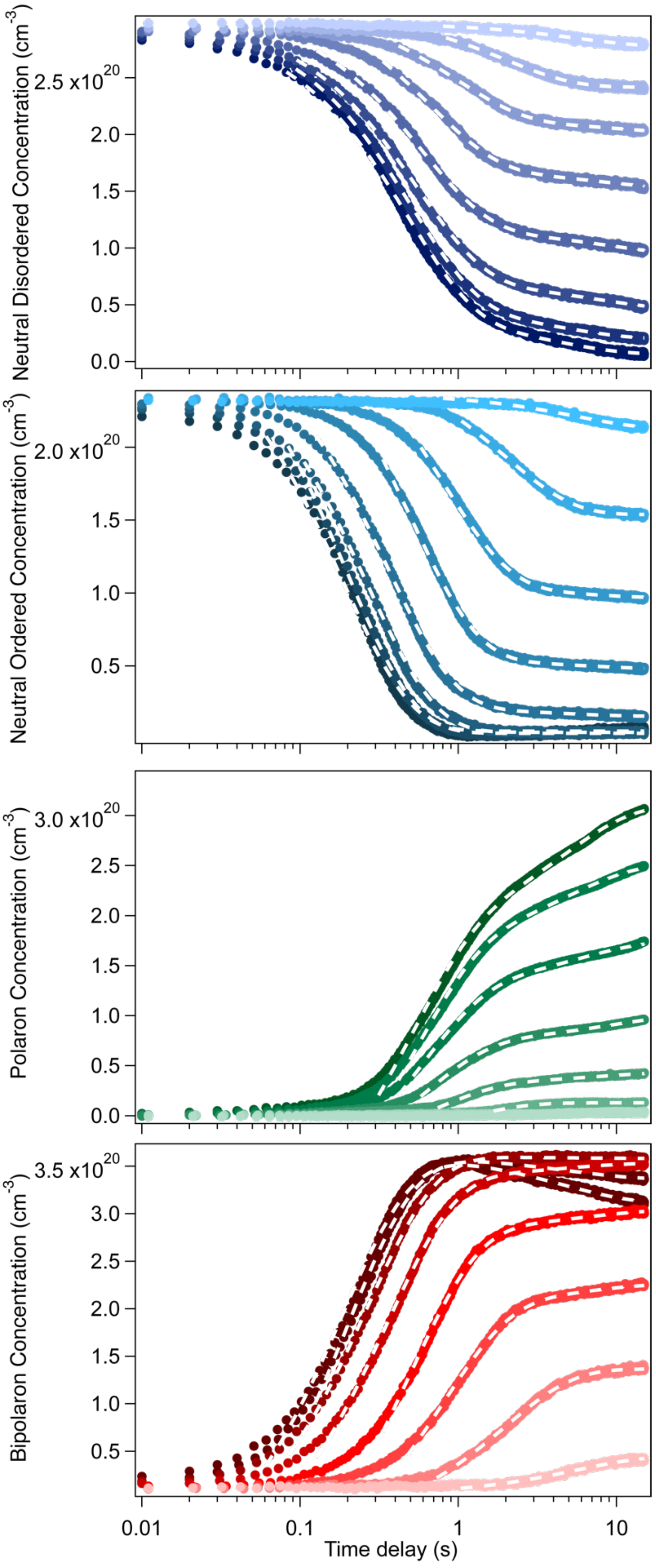

**Supplementary Figure 9**: Same as Figure S7 shown on a logarithmic time axis.

## Supplementary Note 4: Exponential fitting of the dedoping dynamics

The time-resolved spectroelectrochemical dedoping dynamics are shown in Figure S9 for the different redox species at all applied potentials. We find that the curves can generally be analysed with a multiexponential fit and offset according to the equations shown below (D= disordered neutral, O = ordered neutral, P = polaron, B = bipolaron). Depending on the regime, some parameters were held at zero and similar time constants were linked during the fit. The fits are shown in Figure S9 and S10 (log scale) and the parameters summarized in Table S4.

$$D(t) = D_1 exp\left(-t/\tau_{D1}\right) + D_2 exp\left(-t/\tau_{D2}\right) + y_D \qquad (S14)$$

$$O(t) = O_1 exp\left(-t/\tau_{O1}\right) + O_2 exp\left(-t/\tau_{O2}\right) + y_O \qquad (S15)$$

$$P(t) = P_1 exp\left(-t/\tau_{P1}\right) + P_2 exp\left(-t/\tau_{P2}\right) + P_3 exp\left(-t/\tau_{P3}\right) + y_P \qquad (S16)$$

$$B(t) = B_1 exp\left(-t/\tau_{B1}\right) + y_B. \qquad (S17)$$

**Supplementary Table 4**: Parameters obtained by fitting the doping dynamics. Amplitudes are expressed as a % of the total amplitude.

| **V (V)** | **-0.1** | **-0.2** | **-0.3** | **-0.4** | **-0.5** | **-0.6** | **-0.7** | **-0.8** |
|---|---|---|---|---|---|---|---|---|
| $\tau_{D1}(s)$ | | **0.2** (-18%) | **0.2** (-24%) | **0.2** (-40%) | **0.2** (-49%) | **0.2** (-80%) | **0.3** (-86%) | **0.4** (-89%) |
| $\tau_{D2}(s)$ | **0.5** (-100%) | **0.5** (-82%) | **0.5** (-76%) | **0.5** (-60%) | **0.5** (-51%) | **1.0** (-20%) | **1.6** (-14%) | **2.1** (-11%) |
| $y_D$ $(10^{20}cm^{-3})$ | **2.9** (100%) | **2.9** (100%) | **2.9** (100%) | **2.9** (100%) | **2.9** (100%) | **2.9** (100%) | **2.8** (100%) | **2.8** (100%) |
| $\tau_{O1}(s)$ | **0.2** (-84%) | **0.2** (-81%) | **0.2** (-71%) | **0.2** (-70%) | **0.3** (-68%) | **0.3** (-74%) | **0.4** (-81%) | **0.5** (-83%) |
| $\tau_{O2}(s)$ | **1.4** (-16%) | **1.0** (-19%) | **0.9** (-29%) | **0.9** (-30%) | **0.9** (-32%) | **1.0** (-26%) | **1.5** (-19%) | **1.9** (-17%) |
| $y_O$ $(10^{20}cm^{-3})$ | **2.3** (100%) | **2.3** (100%) | **2.3** (100%) | **2.3** (100%) | **2.3** (100%) | **2.3** (100%) | **2.3** (100%) | **2.2** (100%) |
| $\tau_{P1}(s)$ | **0.2** (61%) | **0.2** (69%) | **0.2** (65%) | **0.2** (67%) | **0.3** (69%) | **0.3** (78%) | **0.4** (82%) | **0.5** (84%) |
| $\tau_{P2}(s)$ | **1.4** (17%) | **1.0** (23%) | **0.9** (30%) | **0.9** (28%) | **0.9** (26%) | **1.0** (19%) | **1.5** (14%) | **1.9** (12%) |
| $\tau_{P3}(s)$ | - | - | - | - | - | **0.1** (-100%) | **0.1** (-100%) | **0.1** (-100%) |
| $y_P$ $(10^{20}cm^{-3})$ | **0.1** (22%) | **0.1** (8%) | **0.1** (6%) | **0.1** (5%) | **0.2** (5%) | **0.2** (3%) | **0.2** (4%) | **0.2** (4%) |
| $\tau_{B1}(s)$ | - | - | **0.1** (86%) | **0.1** (92%) | **0.1** (95%) | **0.1** (98%) | **0.2** (98%) | **0.2** (97%) |
| $y_B$ $(10^{20}cm^{-3})$ | - | - | **0.02** (14%) | **0.03** (8%) | **0.05** (5%) | **0.04** (2%) | **0.1** (2%) | **0.1** (3%) |

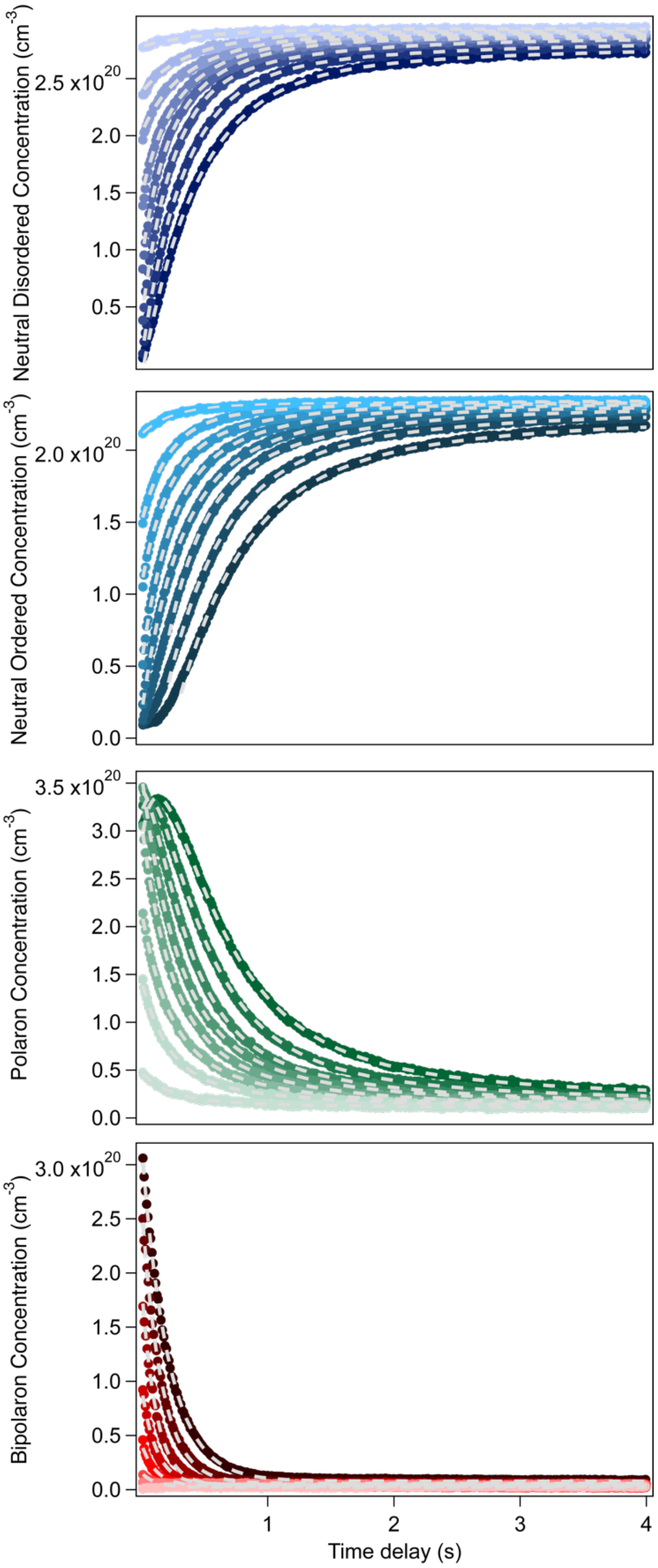

**Supplementary Figure 10**: **Exponential fitting of the time-resolved dedoping dynamics.** Time-resolved densities of the neutral disordered (darker blue), neutral ordered (lighter blue), polaron (green) and bipolaron (red) species obtained from the MCR analysis of the spectroelectrochemistry data when switching the voltage according to Figure S6 during dedoping. From lighter to darker are the different dedoping dynamics to +0.4 V after doping with different applied voltages ranging from -0.1 V to -0.8 V, with steps of -0.1 The fittings are shown as dashed white lines.

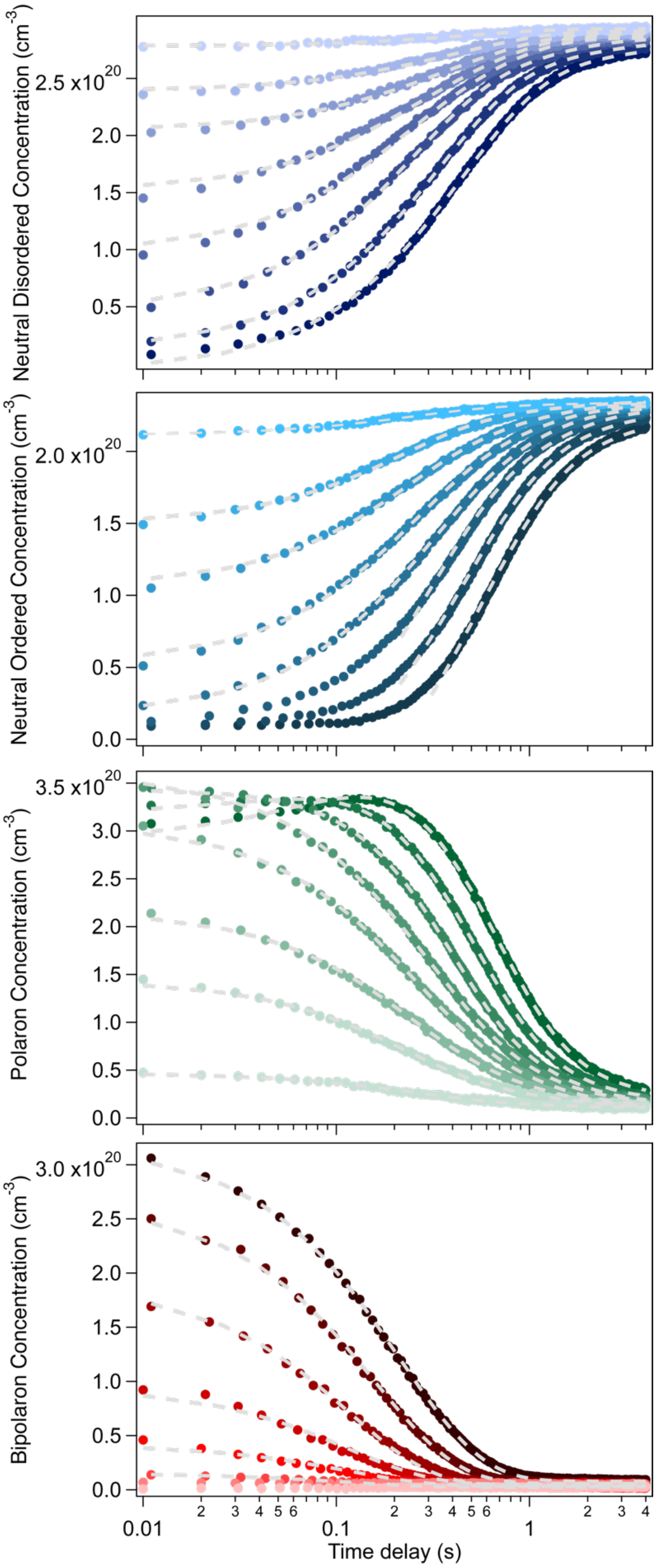

**Supplementary Figure 11**: Same as Figure S9 shown on a logarithmic time axis.

**Supplementary Note 5: Kinetic modelling of the doping proces**

We propose different reaction mechanisms for the doping in each voltage regime, as described by the rate equations below:

**Regime I**

$$N_{dis} \rightleftarrows P_{dis}$$
$$N_{ord} \rightleftarrows P_{ord}$$

Rate constants: $k_{N,dis\rightarrow P,dis}$, $k_{P,dis\rightarrow N,dis}$, $k_{N,ord\rightarrow P,ord}$, $k_{P,ord\rightarrow N,ord}$

Rate equations:

$$\frac{d[N_{dis}]}{dt} = -[N_{dis}]\,k_{N,dis\rightarrow P,dis} + [P_{dis}]\,k_{P,dis\rightarrow N,dis}$$

$$\frac{d[P_{dis}]}{dt} = +[N_{dis}]\,k_{N,dis\rightarrow P,dis} - [P_{dis}]\,k_{P,dis\rightarrow N,dis}$$

$$\frac{d[N_{ord}]}{dt} = -[N_{ord}]\,k_{N,ord\rightarrow P,ord} + [P_{ord}]\,k_{P,ord\rightarrow N,ord}$$

$$\frac{d[P_{ord}]}{dt} = +[N_{ord}]\,k_{N,ord\rightarrow P,ord} - [P_{ord}]\,k_{P,ord\rightarrow N,ord}$$

**Regime II**

$$N_{dis} \rightleftarrows P_{dis} \rightleftarrows B_{dis}$$
$$N_{ord} \rightleftarrows P_{ord}$$

Rate constants: $k_{N,dis\rightarrow P,dis}$, $k_{P,dis\rightarrow N,dis}$, $k_{P,dis\rightarrow B,dis}$, $k_{B,dis\rightarrow P,dis}$, $k_{N,ord\rightarrow P,ord}$, $k_{P,ord\rightarrow N,ord}$

Rate equations:

$$\frac{d[N_{dis}]}{dt} = -[N_{dis}]\,k_{N,dis\rightarrow P,dis} + [P_{dis}]\,k_{P,dis\rightarrow N,dis}$$

$$\frac{d[P_{dis}]}{dt} = +[N_{dis}]\,k_{N,dis\rightarrow P,dis} - [P_{dis}]\,k_{P,dis\rightarrow N,dis} - [P_{dis}]\,k_{P,dis\rightarrow B,dis} + [B_{dis}]\,k_{B,dis\rightarrow P,dis}$$

$$\frac{d[B_{dis}]}{dt} = [P_{dis}]\,k_{P,dis\rightarrow B,dis} - [B_{dis}]\,k_{B,dis\rightarrow P,dis}$$

$$\frac{d[N_{ord}]}{dt} = -[N_{ord}]\,k_{N,ord\rightarrow P,ord} + [P_{ord}]\,k_{P,ord\rightarrow N,ord}$$

$$\frac{d[P_{ord}]}{dt} = +[N_{ord}]\,k_{N,ord\rightarrow P,ord} - [P_{ord}]\,k_{P,ord\rightarrow N,ord}$$

**Regime III**

$$N_{dis} \rightleftarrows P_{dis} \rightleftarrows B_{dis}$$
$$N_{ord} \rightarrow P_{ord} \rightleftarrows B_{ord}$$

Rate constants: $k_{N,dis\to P,dis}$, $k_{P,dis\to N,dis}$, $k_{P,dis\to B,dis}$, $k_{B,dis\to P,dis}$, $k_{N,ord\to P,ord}$, $k_{P,ord\to B,ord}$, $k_{B,ord\to P,ord}$

Rate equations:

$$\frac{d[N_{dis}]}{dt} = -[N_{dis}]\, k_{N,dis\to P,dis} + [P_{dis}]\, k_{P,dis\to N,dis}$$

$$\frac{d[P_{dis}]}{dt} = [N_{dis}]\, k_{N,dis\to P,dis} - [P_{dis}]\, k_{P,dis\to N,dis} - [P_{dis}]\, k_{P,dis\to B,dis} + [B_{dis}]\, k_{B,dis\to P,dis}$$

$$\frac{d[B_{dis}]}{dt} = [P_{dis}]\, k_{P,dis\to B,dis} - [B_{dis}]\, k_{B,dis\to P,dis}$$

$$\frac{d[N_{ord}]}{dt} = -[N_{ord}]\, k_{N,ord\to P,ord}$$

$$\frac{d[P_{ord}]}{dt} = [N_{ord}]\, k_{N,ord\to P,ord} - [P_{ord}]\, k_{P,ord\to B,ord} + [B_{ord}]\, k_{B,ord\to P,ord}$$

$$\frac{d[B_{ord}]}{dt} = [P_{ord}]\, k_{P,ord\to B,ord} - [B_{ord}]\, k_{B,ord\to P,ord}$$

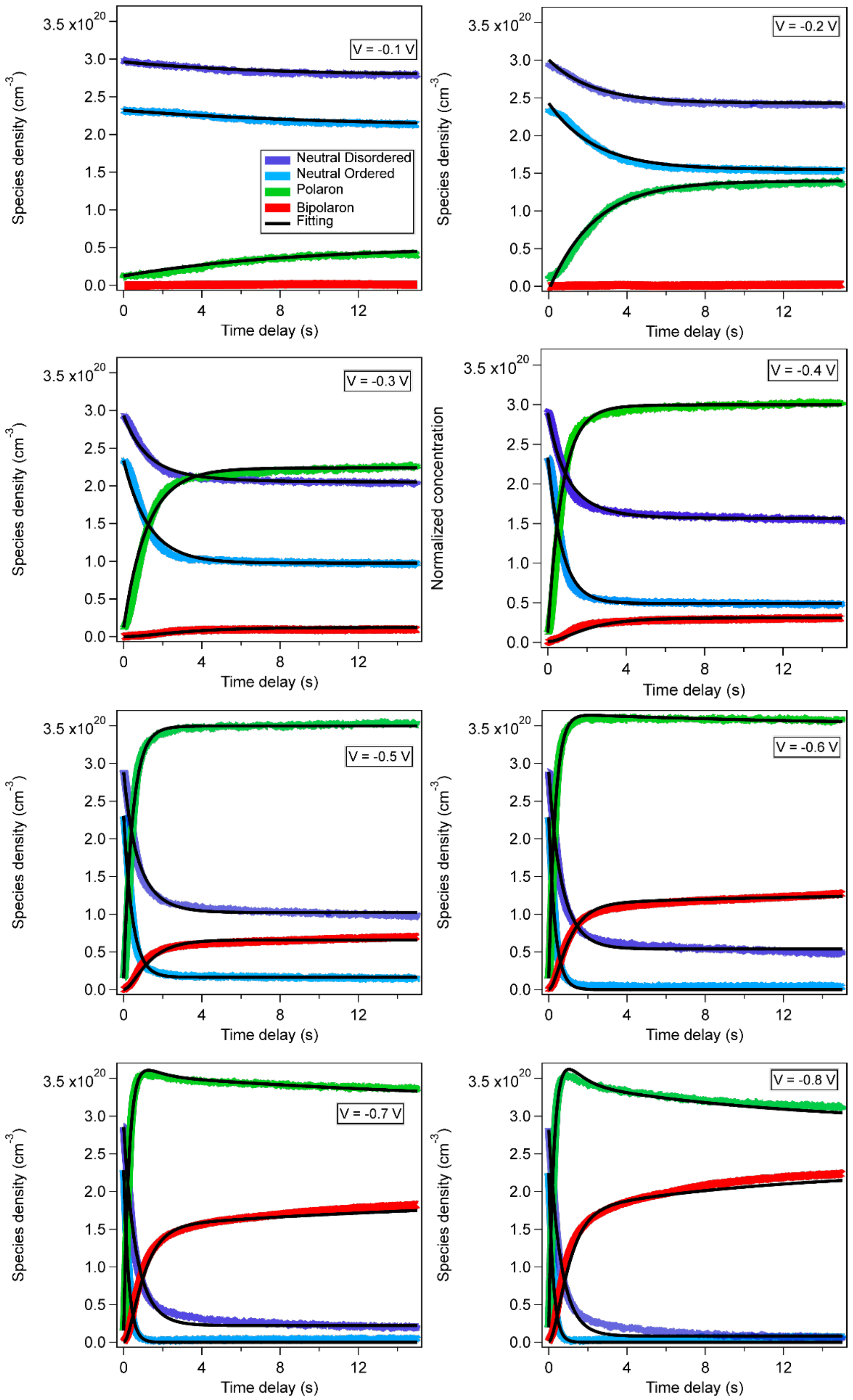

**Supplementary Figure 12**: **Kinetic model fitting (in black) of the species concentrations during the doping from +0.4V to different voltages.** Here, the bipolaron population are divided by 1.37, to keep the total species density constant.

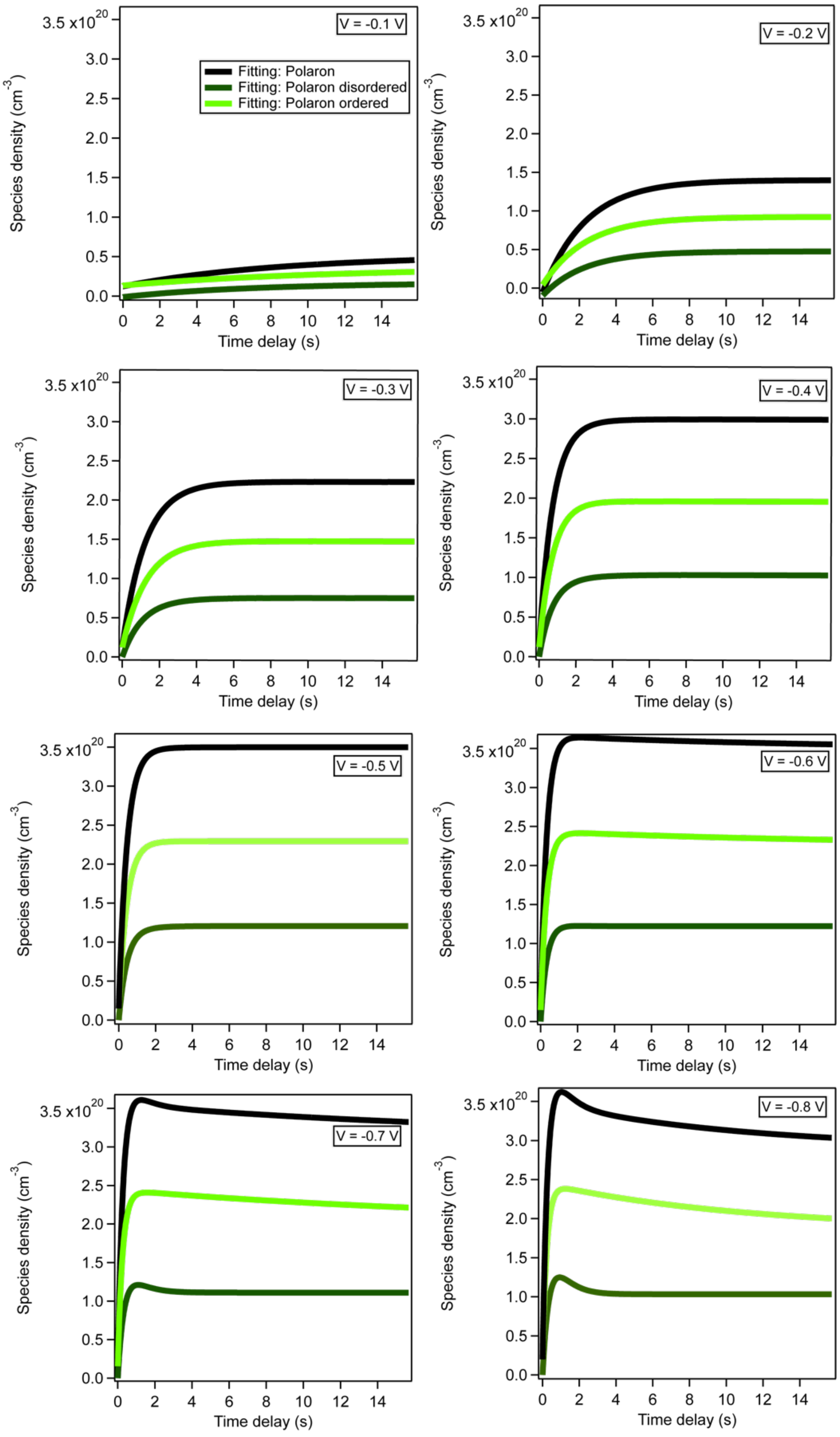

**Supplementary Figure 13: Subpopulations of polarons in different film regions obtained via kinetic modelling.** The total polaron density (black) is the sum of the disordered (dark green) and ordered (light green) subpopulations.

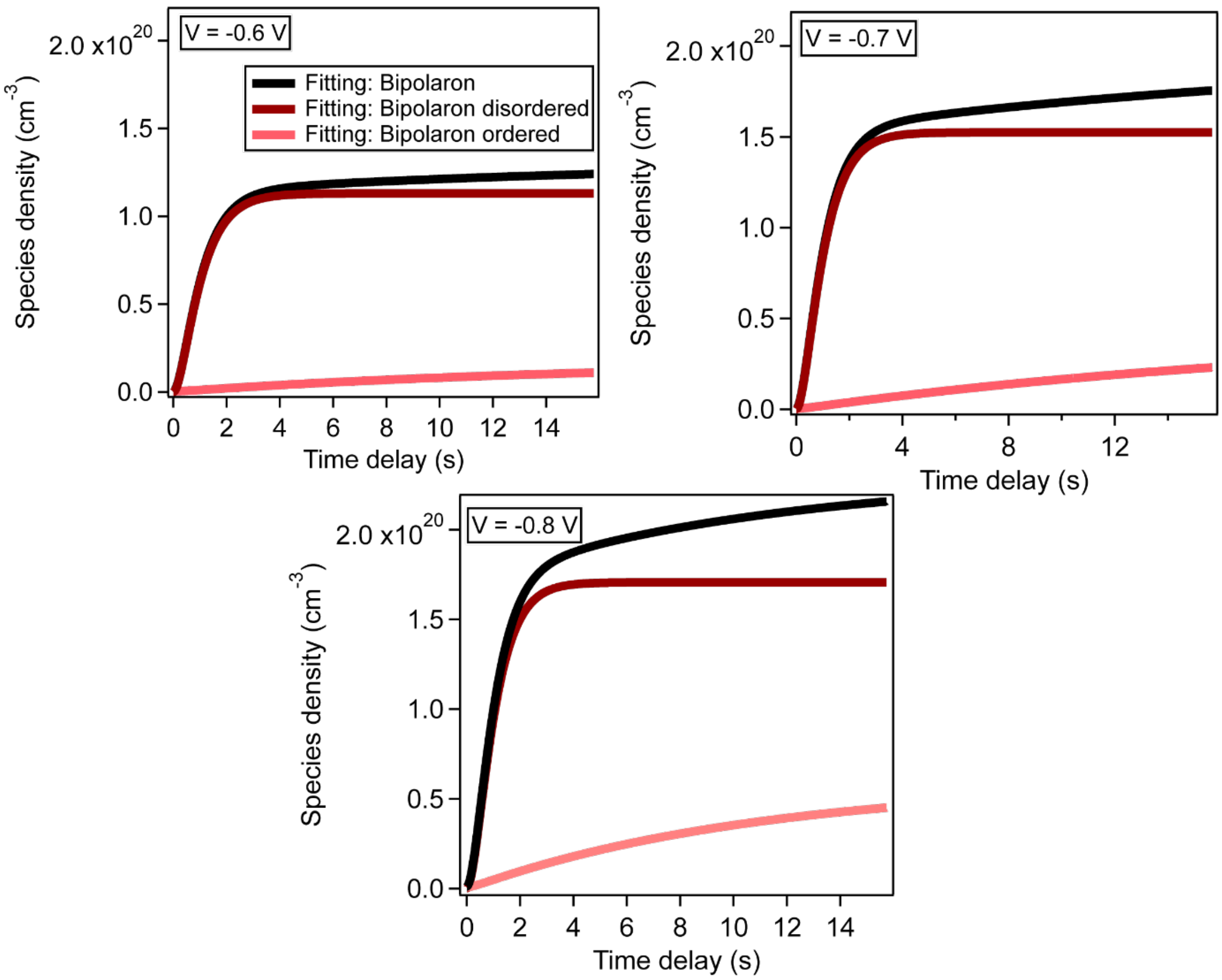

**Supplementary Figure 14: Subpopulations of bipolarons in different film regions obtained via kinetic modelling.** The total bipolaron density (black) is the sum of the disordered (dark red) and ordered (light red) subpopulations. The bipolaron population are divided by 1.37, to keep the total species density constant.

## Supplementary Note 6: Kinetic modelling of the dedoping process

We propose different reaction mechanisms for the dedoping in each voltage regime, as described by the rate equations below:

**Regime I**

$$P_{dis} \rightleftarrows N_{dis}$$
$$P_{ord} \rightleftarrows N_{ord}$$

Rate constants: $k_{N,dis \to P,dis}$, $k_{P,dis \to N,dis}$, $k_{N,ord \to P,ord}$, $k_{P,ord \to N,ord}$

Rate equations:

$$\frac{d[P_{dis}]}{dt} = +[N_{dis}]\,k_{N,dis \to P,dis} - [P_{dis}]\,k_{P,dis \to N,dis}$$

$$\frac{d[N_{dis}]}{dt} = -[N_{dis}]\,k_{N,dis \to P,dis} + [P_{dis}]\,k_{P,dis \to N,dis}$$

$$\frac{d[P_{ord}]}{dt} = +[N_{ord}]\,k_{N,ord \to P,ord} - [P_{ord}]\,k_{P,ord \to N,ord}$$

$$\frac{d[N_{ord}]}{dt} = -[N_{ord}]\,k_{N,ord \to P,ord} + [P_{ord}]\,k_{P,ord \to N,ord}$$

**Regime II**

$$B_{dis} \rightarrow P_{dis} \rightleftarrows N_{dis}$$
$$P_{ord} \rightleftarrows N_{ord}$$

Rate constants: $k_{N,dis \to P,dis}$, $k_{P,dis \to N,dis}$, $k_{B,dis \to P,dis}$, $k_{N,ord \to P,ord}$, $k_{P,ord \to N,ord}$

Rate equations:

$$\frac{d[B_{dis}]}{dt} = -[B_{dis}]\,k_{B,dis \to P,dis}$$

$$\frac{d[P_{dis}]}{dt} = [B_{dis}]\,k_{B,dis \to P,dis} - [P_{dis}]\,k_{P,dis \to N,dis} + [N_{dis}]\,k_{N,dis \to P,dis}$$

$$\frac{d[N_{dis}]}{dt} = [P_{dis}]\,k_{P,dis \to N,dis} - [N_{dis}]\,k_{N,dis \to P,dis}$$

$$\frac{d[P_{ord}]}{dt} = [N_{ord}]\,k_{N,ord \to P,ord} - [P_{ord}]\,k_{P,ord \to N,ord}$$

$$\frac{d[N_{ord}]}{dt} = -[N_{ord}]\,k_{N,ord \to P,ord} + [P_{ord}]\,k_{P,ord \to N,ord}$$

**Regime III**

$$B_{dis} \rightarrow P_{dis} \rightleftarrows N_{dis}$$
$$B_{ord} \rightarrow P_{ord} \rightleftarrows N_{ord}$$

Rate constants: $k_{N,dis \rightarrow P,dis}$, $k_{P,dis \rightarrow N,dis}$, $k_{P,dis \rightarrow B,dis}$, $k_{B,dis \rightarrow P,dis}$, $k_{N,ord \rightarrow P,ord}$, $k_{B,ord \rightarrow P,ord}$

Rate equations:

$$\frac{d[B_{dis}]}{dt} = -[B_{dis}]\, k_{B,dis \rightarrow P,dis}$$

$$\frac{d[P_{dis}]}{dt} = [B_{dis}]\, k_{B,dis \rightarrow P,dis} - [P_{dis}]\, k_{P,dis \rightarrow N,dis} + [N_{dis}]\, k_{N,dis \rightarrow P,dis}$$

$$\frac{d[N_{dis}]}{dt} = [P_{dis}]\, k_{P,dis \rightarrow N,dis} - [N_{dis}]\, k_{N,dis \rightarrow P,dis}$$

$$\frac{d[B_{ord}]}{dt} = -[B_{ord}]\, k_{B,ord \rightarrow P,ord}$$

$$\frac{d[P_{ord}]}{dt} = +[B_{ord}]\, k_{B,ord \rightarrow P,ord} + [N_{ord}]\, k_{N,ord \rightarrow P,ord} - [P_{ord}]\, k_{P,ord \rightarrow N,ord}$$

$$\frac{d[N_{ord}]}{dt} = [P_{ord}]\, k_{P,ord \rightarrow N,ord} - [N_{ord}]\, k_{N,ord \rightarrow P,ord}$$

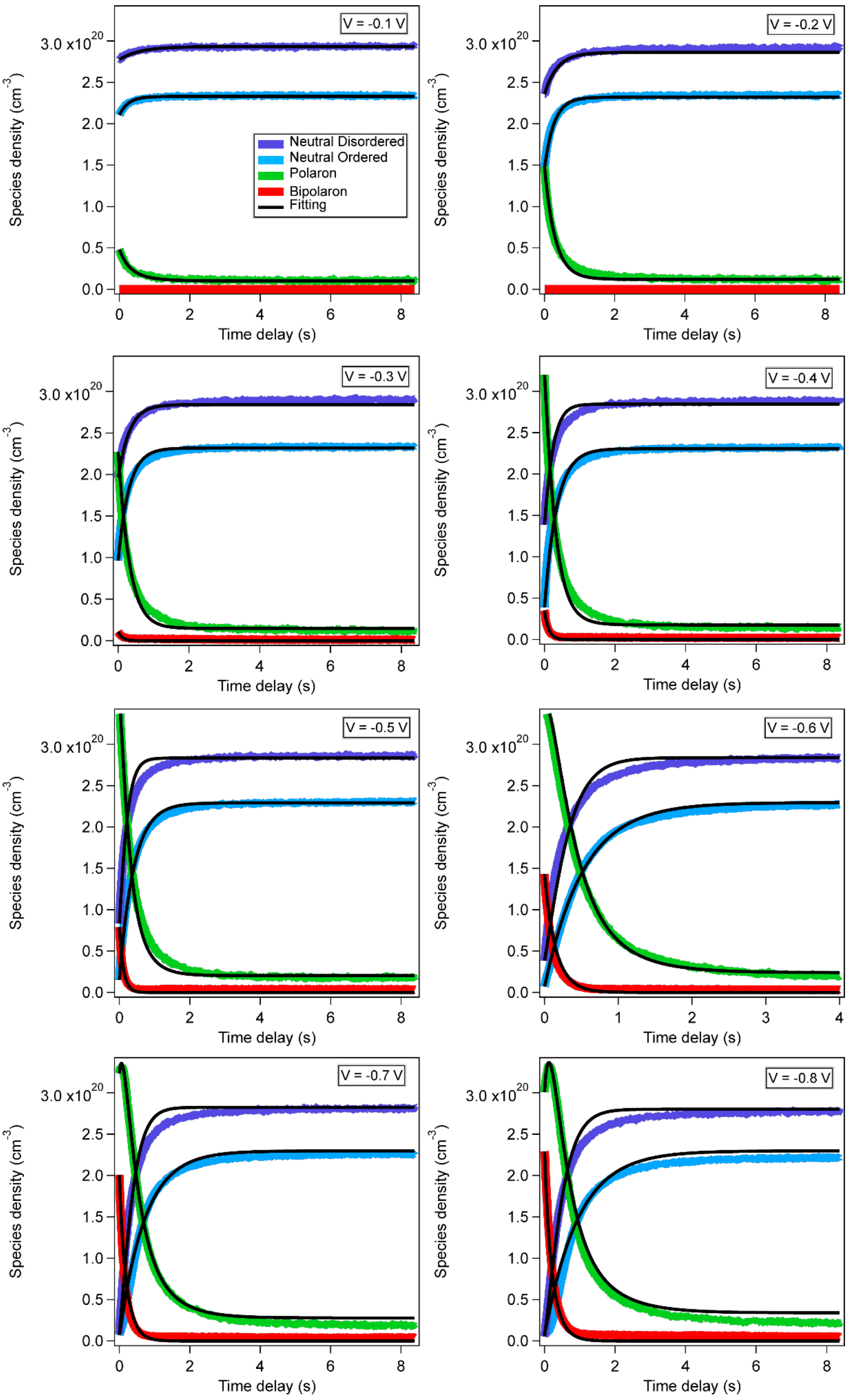

**Supplementary Figure 15: Kinetic model fitting (in black) of the species concentrations during the dedoping from different voltages to +0.4V.**

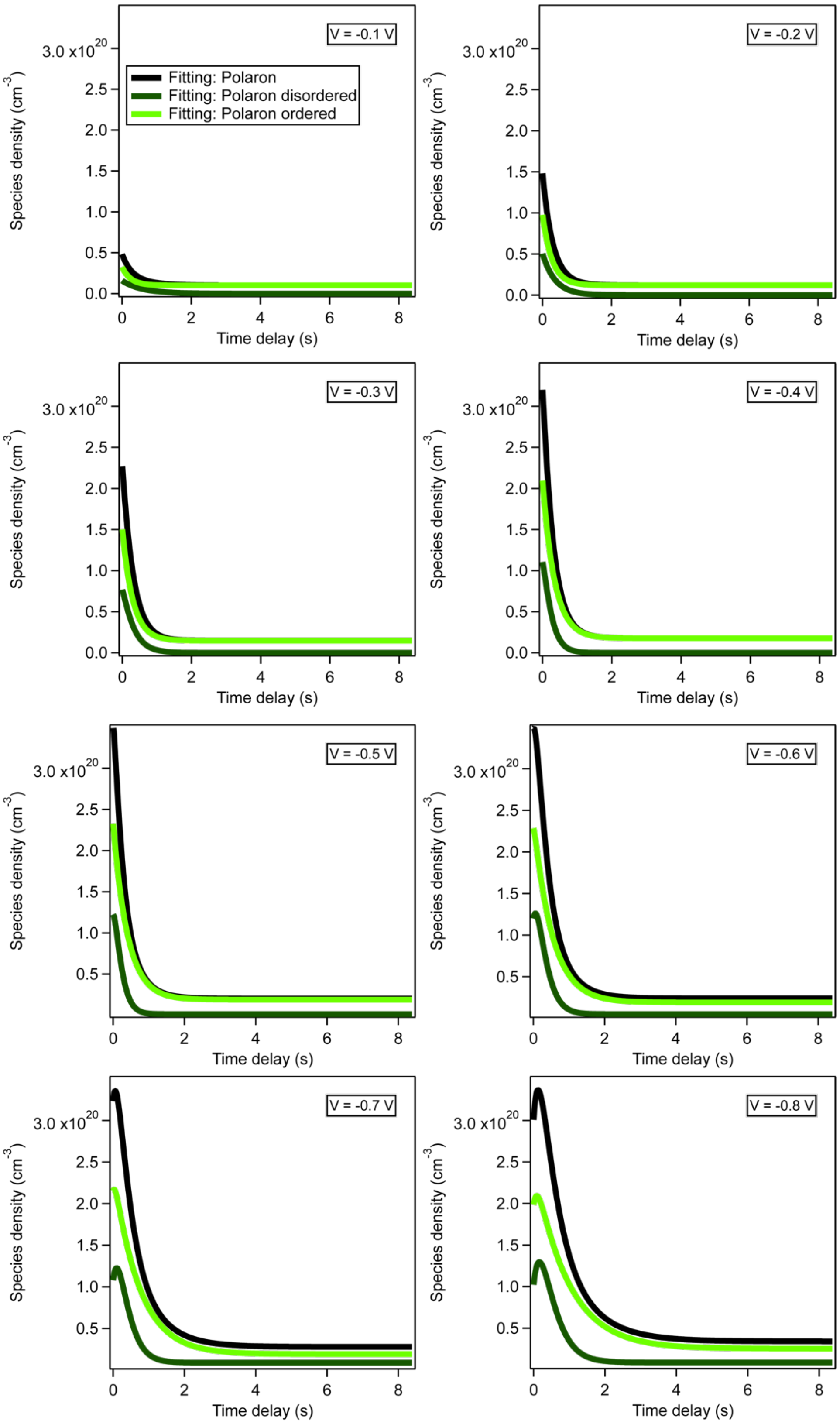

**Supplementary Figure 16: Subpopulations of polarons in different film regions during the dedoping obtained via kinetic modelling.** The total polaron density (black) is the sum of the disordered (dark green) and ordered (light green) subpopulations.

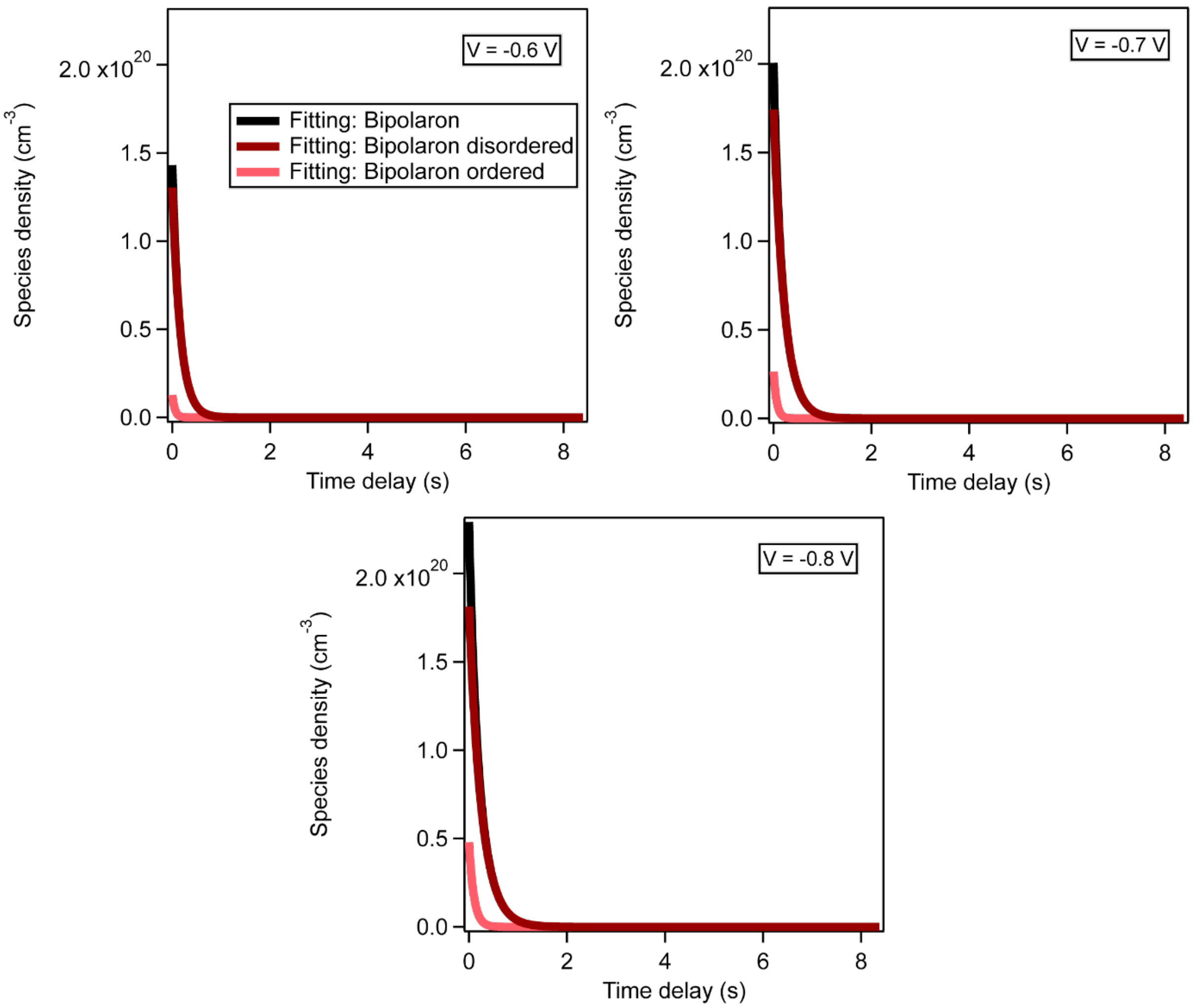

**Figure S17: Subpopulations of bipolarons in different film regions during the dedoping obtained via kinetic modelling.** The total bipolarons density (black) is the sum of the disordered (dark red) and ordered (light red) subpopulations.

.

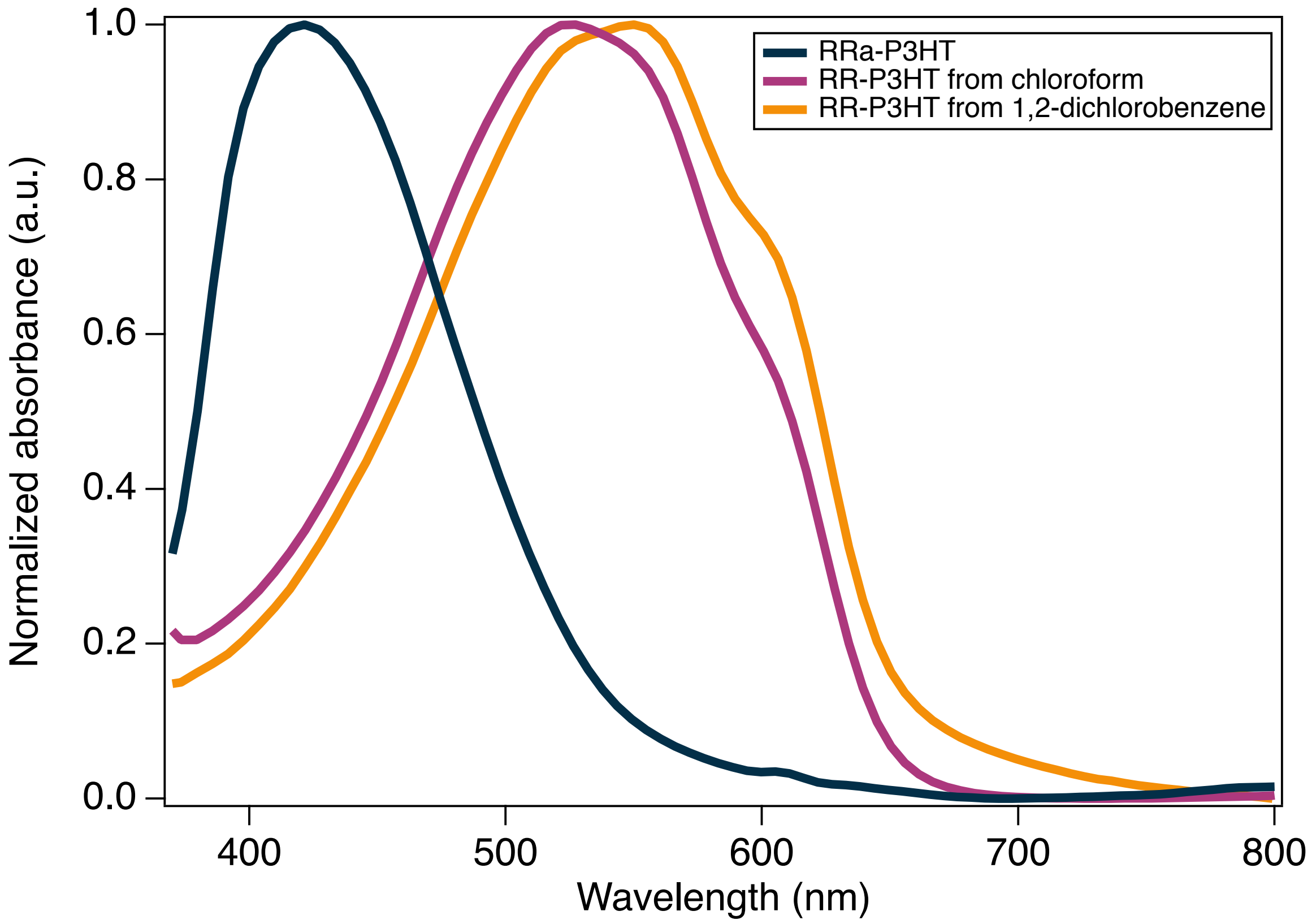

**Supplementary Figure 18: Absorbance spectra of P3HT films with different morphology**. Comparison between the normalized absorbance spectra of the dry RRa-P3HT and RR-P3HT films with different degrees of crystallinity (cast from chloroform or 1,2-dichlorobenzene).

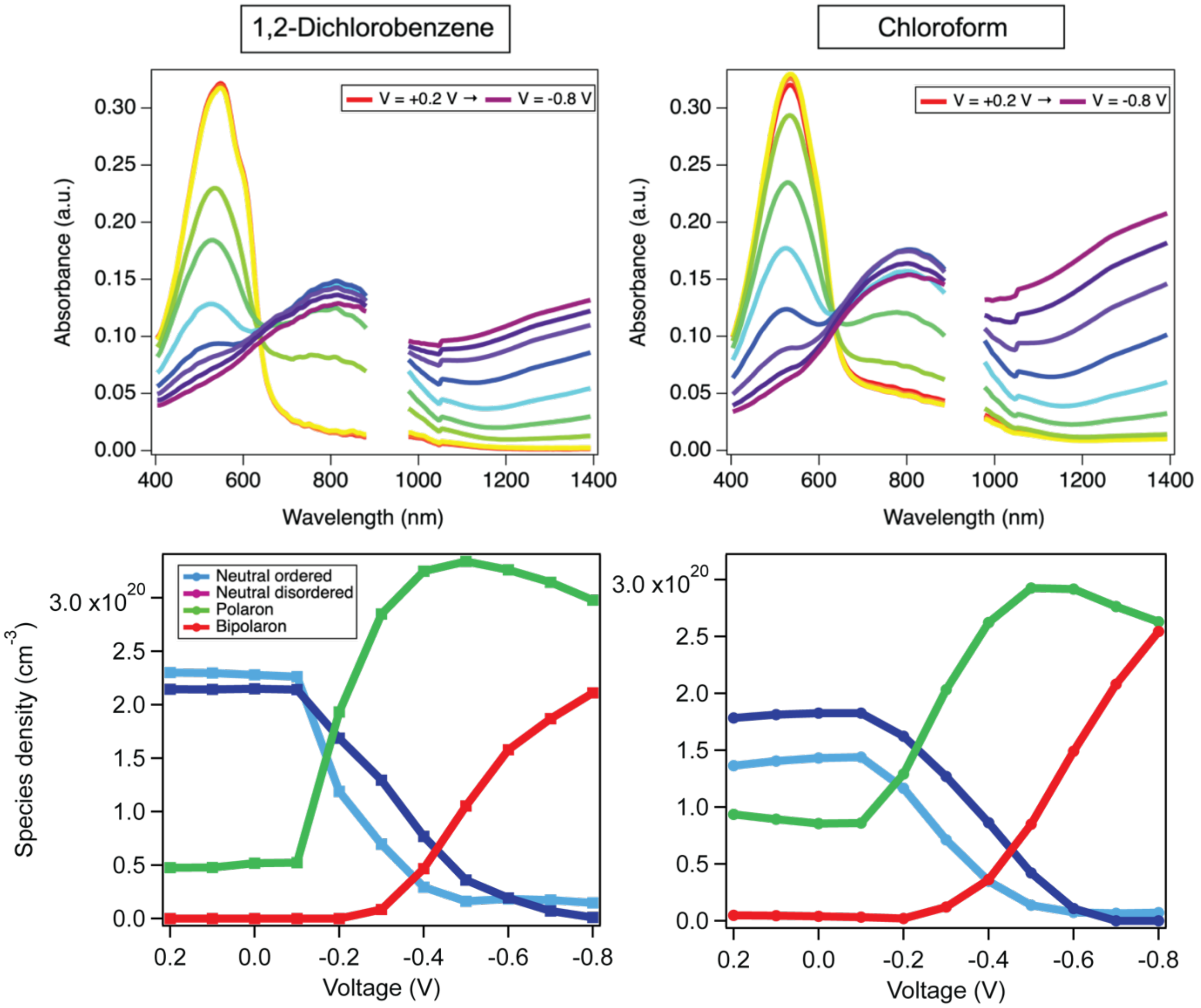

**Supplementary Figure 19**: **Doping of RR-P3HT films with different degrees of order**. The top panels show the absorbance spectra of RR-P3HT films cast from 1,2-dichlorobenze (on the left) and chloroform (on the right), at different electrochemical potentials. The films are initially slightly doped, because not enough time was waited after initially cycling the device and performing these measurements. This causes the initial polaron population. It can also be seen that even though the films have similar thickness and similar absorbance at the neutral and polaron bands, the bipolarons rise much higher in the chloroform films. The lower panels show the species densities at different voltages after using MCR to decompose the spectra. The lower order (in the ratio of neutral species) and more important bipolaron formation in the chloroform-cast film is confirmed.

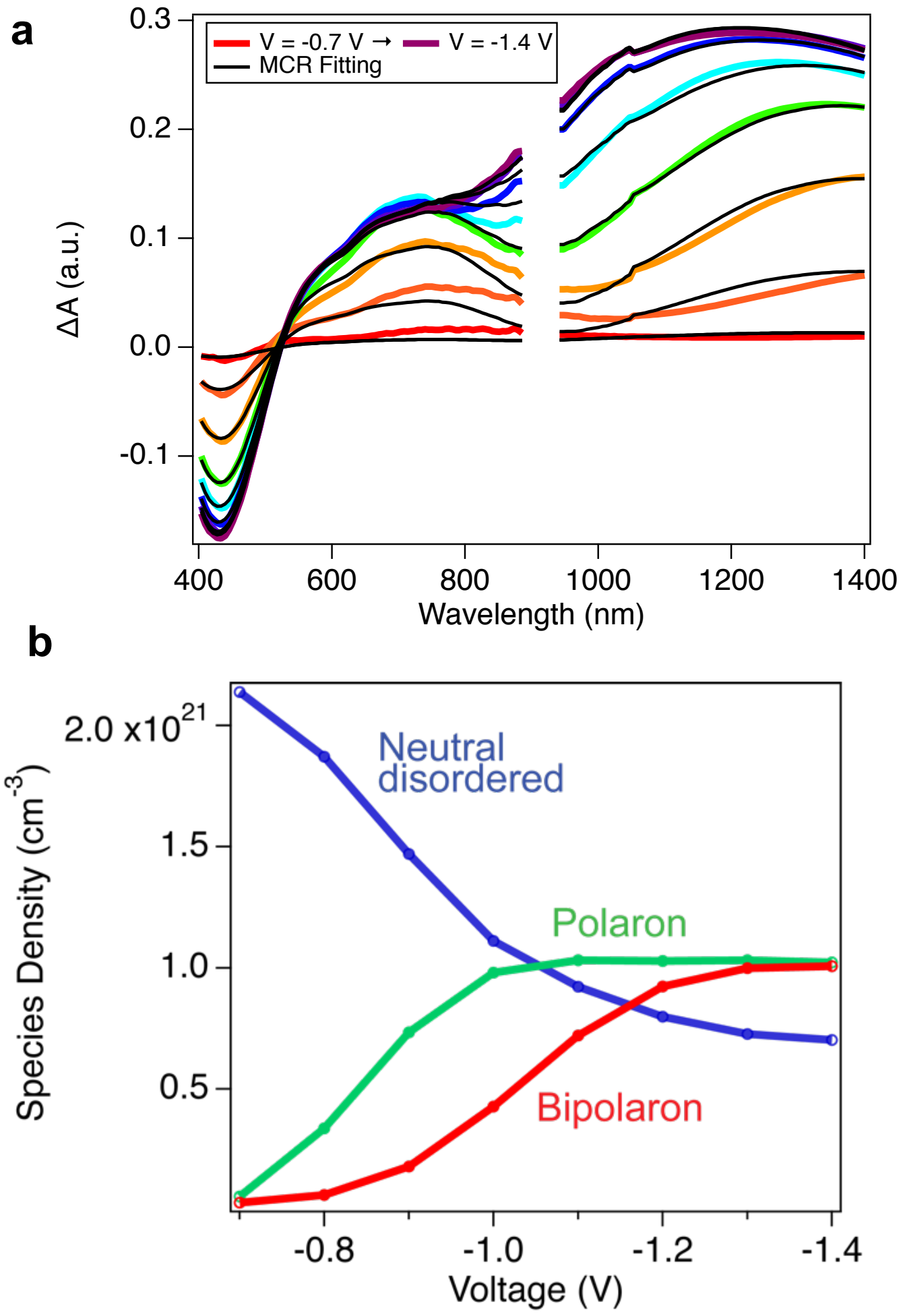

**Supplementary Figure 20**: **Spectroelectrochemical measurements for the RRa-P3HT film**. a) Differential steady-state absorbance of the RRa-P3HT film doped in a 0.1 M $KPF_6$ acetonitrile electrolyte. The fit resulting from the MCR analysis with three components is shown in black. b) Species density for the different voltages.

## Supplementary Note 7: Exponential fitting of the dynamics in RRa-P3HT

The time-resolved species density during the doping from 0 to -1.2 V and dedoping from -1.2 to 0 V were analysed using bi-exponential functions with an offset, as shown below. The obtained rate constants are summarized in Table S5 and the fitting results are shown in Figure S19.

$$D(t) = D_1 exp\left(-t/_{\tau_{D1}}\right) + D_2 exp\left(-t/_{\tau_{D2}}\right) + y_D \qquad S18$$

$$P(t) = P_1 exp\left(-t/_{\tau_{P1}}\right) + P_2 exp\left(-t/_{\tau_{P2}}\right) + y_P \qquad S19$$

$$B(t) = B_1 exp\left(-t/_{\tau_{B1}}\right) + B_2 exp\left(-t/_{\tau_{B2}}\right) + y_B. \qquad S20$$

**Table S5**: Parameters obtained by fitting the doping and dedoping dynamics in RRa-P3HT. Amplitudes are expressed as a % of the total amplitude.

| | ***Doping*** | ***Dedoping*** |
|---|---|---|
| $\tau_{D1}(s)$ | **0.2** (18%) | **0.1** (-95%) |
| $\tau_{D2}(s)$ | **2.3** (48%) | **2.5** (-5%) |
| $y_D(10^{20}cm^{-3})$ | **7.8** (34%) | **20.8** (100%) |
| $\tau_{P1}(s)$ | **0.2** (-95%) | **0.1** (90%) |
| $\tau_{P2}(s)$ | **1.1** (-5%) | **0.7** (5%) |
| $y_P\ (10^{20}cm^{-3})$ | **10.1** (100%) | **1.4** (5%) |
| $\tau_{B1}(s)$ | **0.3** (-47%) | **0.1** (93%) |
| $\tau_{B2}(s)$ | **2.3** (-53%) | **-** |
| $y_B\ (10^{20}cm^{-3})$ | **8.7** (100%) | **0.6** (7%) |

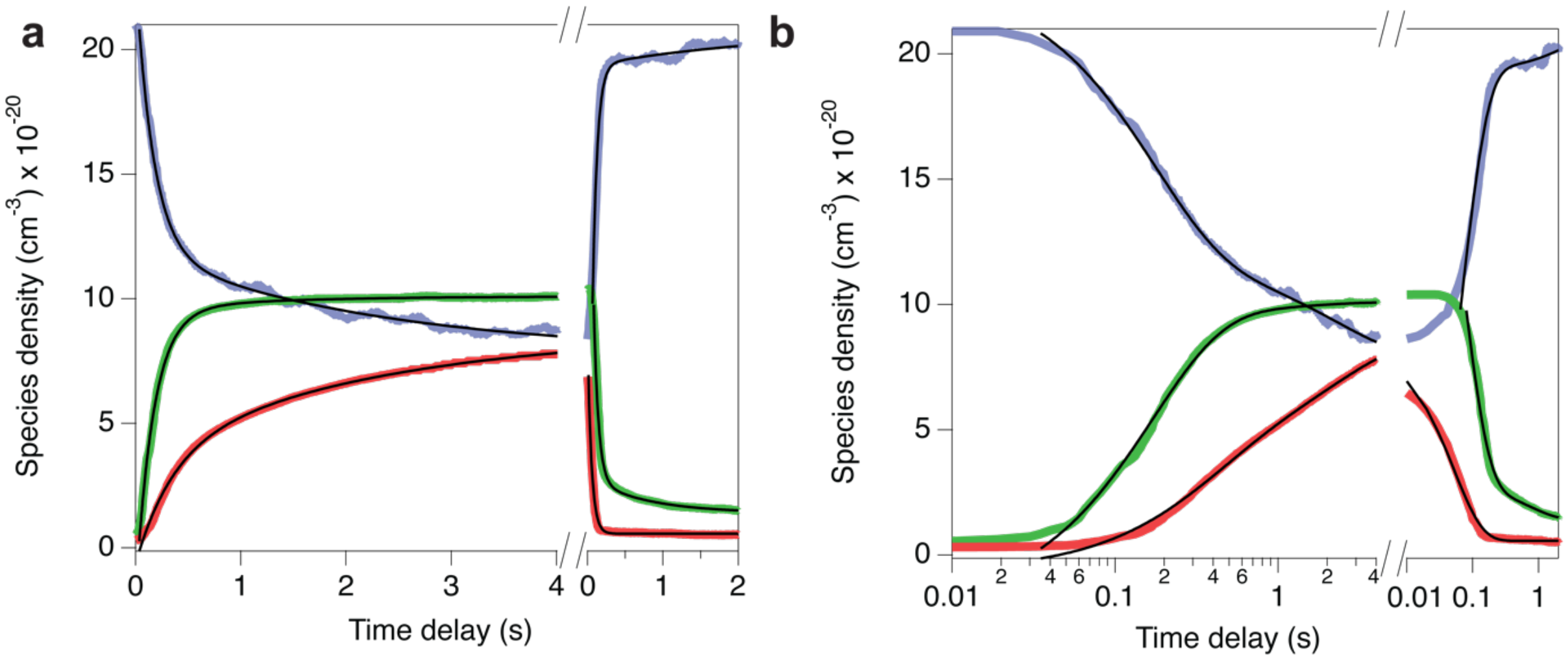

**Supplementary Figure 21**: **Exponential fitting of the time-resolved dynamics in RRa-P3HT**. Exponential fitting (in black) of the species concentrations during the doping (from 0 to -1.2 V) and dedoping (from -1.2 V to 0 V) of RRa-P3HT film. Neutral disordered species (dark blue), polarons (green) and bipolarons (red) are shown, in a) on a linear time axis and in b) on a logarithmic time axis.

**Supplementary Note 8 Kinetic modelling of the doping and dedoping processes in RRa-P3HT**

We propose the following reaction mechanisms for the doping step from 0 to -1.2 V of the RRa-P3HT film, described by the rate equations below:

$$N_{dis} \rightleftarrows P_{dis} \rightleftarrows B_{dis}$$

Rate constants: $k_{N,dis \to P,dis}$, $k_{P,dis \to N,dis}$, $k_{P,dis \to B,dis}$, $k_{B,dis \to P,dis}$

Rate equations:

$$\frac{d[N_{dis}]}{dt} = -[N_{dis}]\,k_{N,dis \to P,dis} + [P_{dis}]\,k_{P,dis \to N,dis}$$

$$\frac{d[P_{dis}]}{dt} = +[N_{dis}]\,k_{N,dis \to P,dis} - [P_{dis}]\,k_{P,dis \to N,dis} - [P_{dis}]\,k_{P,dis \to B,dis} + [B_{dis}]\,k_{B,dis \to P,dis}$$

$$\frac{d[B_{dis}]}{dt} = [P_{dis}]\,k_{P,dis \to B,dis} - [B_{dis}]\,k_{B,dis \to P,dis}$$

For the dedoping step from -1.2 to 0 V, we propose the following:

$$B_{dis} \rightleftarrows P_{dis} \rightleftarrows N_{dis}$$

Rate constants: $k_{N,dis \to P,dis}$, $k_{P,dis \to N,dis}$, $k_{B,dis \to P,dis}$, $k_{P,dis \to B,dis}$

Rate equations:

$$\frac{d[B_{dis}]}{dt} = -[B_{dis}]\,k_{B,dis \to P,dis} + [P_{dis}]\,k_{P,dis \to B,dis}$$

$$\frac{d[P_{dis}]}{dt} = [B_{dis}]\,k_{B,dis \to P,dis} - [P_{dis}]\,k_{P,dis \to N,dis} - [P_{dis}]\,k_{P,dis \to B,dis} + [N_{dis}]\,k_{N,dis \to P,dis}$$

$$\frac{d[N_{dis}]}{dt} = [P_{dis}]\,k_{P,dis \to N,dis} - [N_{dis}]\,k_{N,dis \to P,dis}$$

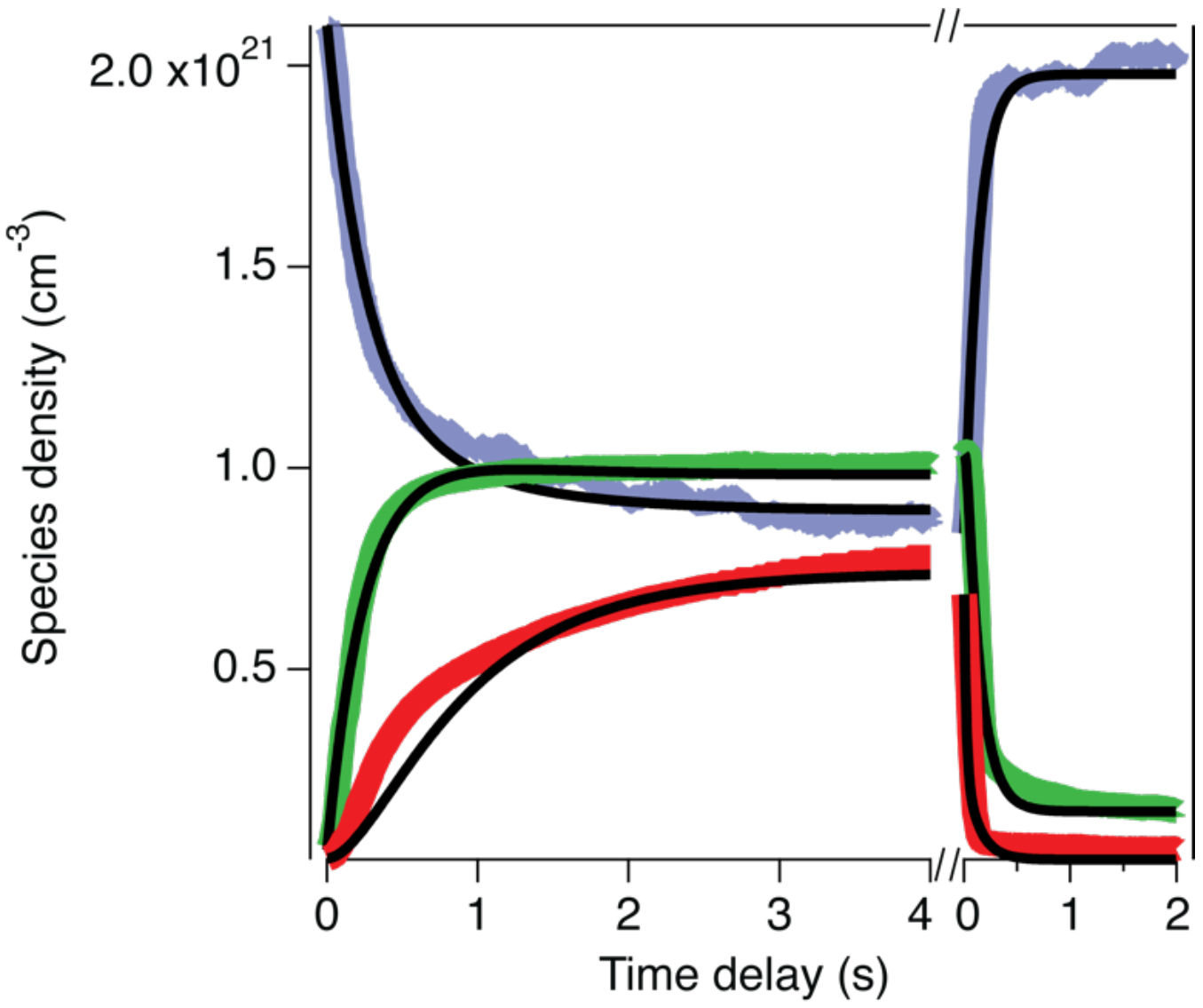

**Supplementary Figure 22**: **Kinetic modelling of the doping and dedoping in RRa-P3HT.** Kinetic model fitting (in black) of the species concentrations during the doping (from 0 to -1.2 V) and dedoping (from -1.2 V to 0 V) of RRa-P3HT film. Neutral disordered species (dark blue), polarons (green) and bipolarons (red) are shown.

**Supplementary Table 6:** Rate constants (k in $s^{-1}$) and equilibrium constants (K, unitless) for the electrochemical doping (from 0 V to - 1.2 V) and dedoping (from -1.2 V to 0 V) of RRa-P3HT, obtained via kinetic modelling. The equilibrium constants are defined as the forward rate constant divided by the backward rate constant.

| *Doping* | | *Dedoping* | |
|---|---|---|---|
| $k_{N,dis \rightarrow P}$ | 1.94 | $k_{B \rightarrow \mathrm{P}}$ | 46.50 |
| $K_{N,dis \rightleftharpoons P}$ | 1.1 | $\mathrm{K}_{\mathrm{B} \rightleftharpoons \mathrm{P}}$ | 14.0 |
| $k_{P \rightarrow B}$ | 0.36 | $k_{P \rightarrow \mathrm{N,dis}}$ | 7.62 |
| $K_{P \rightleftharpoons B}$ | 0.3 | $\mathrm{K}_{P \rightleftharpoons \mathrm{N},dis}$ | 13.6 |

**Supporting References**